\pdfoutput=1
\documentclass[aps, prd, 10pt, twocolumn ,tightenlines,nofootinbib,superscriptaddress,floatfix,preprintnumbers]{revtex4-1}

\usepackage{epsfig}
\usepackage{amsmath}
\usepackage{hyperref}
\usepackage{amssymb}
\usepackage{color}
\newcommand{\be}{\begin{equation}}
\newcommand{\ee}{\end{equation}}
\newcommand{\ba}{\begin{eqnarray}}
\newcommand{\ea}{\end{eqnarray}}
\newcommand{\bd}{\begin{displaymath}}
\newcommand{\ed}{\end{displaymath}}

\newcommand{\commentout}[1]{{}}

\def\rootg{\sqrt{-g}}

\def\DL{\mathcal{D}_L}
\def\W{\mathfrak{w}}
\def\z{\tilde{z}}
\def\u{\tilde{u}}

\begin{document}

  \title{{\bf A shear spectral sum rule in a non-conformal gravity dual}}
  
  \author{Todd Springer}
  \email{springer@physics.mcgill.ca}

  \affiliation{Department of Physics, McGill University, Montreal, Quebec, H3A2T8, Canada}
  \date{Jan. 17, 2011}

  \author{Charles Gale}
  \email{gale@physics.mcgill.ca}
  \affiliation{Department of Physics, McGill University, Montreal, Quebec, H3A2T8, Canada}

  \author{Sangyong Jeon}
  \email{jeon@physics.mcgill.ca}
  \affiliation{Department of Physics, McGill University, Montreal, Quebec, H3A2T8, Canada}

  \author{Su Houng Lee}
  \email{suhoung@phya.yonsei.ac.kr}
  \affiliation{Institute of Physics and Applied Physics, Yonsei University, Seoul 120-749, Korea}
  \affiliation{Yukawa Institute for Theoretical Physics, Kyoto University, Kyoto 606-8502, Japan}

  
  \preprint{YITP-10-50}
  
  \begin{abstract}
  A sum rule which relates a stress-energy tensor correlator to
  thermodynamic functions is examined within the context of a simple
  non-conformal gravity dual.  Such a sum rule was previously derived
  using AdS/CFT for conformal $\mathcal{N} = 4$ Supersymmetric
  Yang-Mills theory, but we show that it does not generalize to the
  non-conformal theory under consideration.  We provide a generalized
  sum rule and numerically verify its validity.  A useful byproduct of
  the calculation is the computation of the spectral density in a
  strongly coupled non-conformal theory.  Qualitative features of the
  spectral densities and implications for lattice measurements of
  transport coefficients are discussed.

  \end{abstract}
\maketitle

\section{Introduction}
Sum rules are powerful tools useful in the exploration of
nonperturbative phenomena.  The use of sum rules in this way dates
back nearly three decades now \cite{Shifman1979385}; but recently
there has been some interest in applying these tools to the strongly
coupled plasma created at the Relativistic Heavy Ion Collider (RHIC)
\cite{Adams:2005dq,Adcox:2004mh,Back:2004je,Arsene:2004fa}.  In
\cite{Karsch:2007jc,Kharzeev:2007wb}, the authors used the low energy
theorems of \cite{Ellis:1998kj} to write down a sum rule which relates
an integral over the spectral density to thermodynamic quantities.  In
general such a sum rule provides constraints on the spectral function,
which could itself be used to extract transport coefficients (via
Kubo's formulas).  Constraints or some knowledge of the functional
form of the spectral density are generally needed in order to have any
hope of extracting transport coefficients from the lattice.  The
authors of \cite{Karsch:2007jc,Kharzeev:2007wb} were able to argue for
qualitative features of the bulk viscosity of the quark-gluon plasma
(QGP) by combining their sum rule with both an ansatz for the spectral
density and lattice data.  This approach was later criticized in
\cite{Moore:2008ws}, and later some corrections and clarifications
were added in \cite{Romatschke:2009ng}.  This latter work also derived
several other sum rules using Kramers-Kroenig relations.  Subsequent
works have derived additional sum rules, and examined the applications
of such sum rules to lattice computations
\cite{Baier:2009zy,Meyer:2010ii, Meyer:2010gu}.  One such sum rule derived in
\cite{Romatschke:2009ng} was derived using AdS/CFT and was found to be
applicable to $\mathcal{N} = 4$ supersymmetric Yang-Mills (SYM)
theory.  It is the aim of this paper to examine this sum rule within the context of a non-conformal gravity
dual theory using the tools of the AdS/CFT correspondence
\cite{Maldacena:1997re, Witten:1998qj, Gubser:1998bc, Son:2002sd}.

The sum rule in question is \cite{Romatschke:2009ng}
\begin{equation}
  \frac{2}{5} \varepsilon 
  = \frac{2}{\pi} \int_0^{\infty} \frac{dw}{w} \left[\rho^{\rm shear}(w) - \rho^{\rm shear}_{\rm T=0}(w) \right] 
  , \label{shearsumrule}
\end{equation}
Here $\varepsilon$ is the energy density, and
$\rho$ is the spectral density
\begin{eqnarray}
  \rho^{\rm shear}(w) \equiv -\mbox{Im }G_R^{\rm shear}(w).
\end{eqnarray}
The retarded Green's function in the ``tensor'' channel is defined as:
\begin{equation}
  G_R^{\rm shear}(w) \equiv -i \int d^4x\, e^{i w t} \left< \left[T^{xy} (x),T^{xy} (0) \right] \right> \theta(t)
\end{equation}
and the subscript $T = 0$ means the quantity of interest is evaluated
in the limit of zero temperature.

Both sides of the sum rule can be computed using AdS/CFT techniques.
The left side depends only on thermodynamic quantities, which are
easily evaluatable for the theory of interest. In order to evaluate
the right hand side, one needs to compute the spectral density $\rho$
as a function of $w$.  In AdS/CFT, the differential equations
necessary to compute spectral densities are often difficult to solve
analytically (though in some cases analytical results have been given
in the literature \cite{Myers:2007we}).  In this work we will solve
the differential equations numerically, and hence our verification of
the sum rule will be numerical in nature. 
\commentout{
In order to evaluate
the right hand side, one needs to compute the spectral density $\rho$
as a function of $w$ numerically.  Any verification of the sum rule
will have to be numerical in nature.}

In \cite{Romatschke:2009ng}, the authors checked that the left and right
sides of the sum rule (\ref{shearsumrule}) are in agreement within the
context of the (conformal) $\mathcal{N} = 4$ SYM
theory.  However, the authors then state that the sum rule should hold
for \emph{any} Einstein gravity dual. 
As shown below, this is actually not the case.
We have evaluated the
left and right sides of the sum rule (\ref{shearsumrule}) in a
particular non-conformal gravity dual theory and find that the left
side is not, in general, equal to the right side.  In fact, one should
expect that the sum rule should be corrected as
\begin{eqnarray}
\frac{2}{5}\varepsilon +
  F(\varepsilon,P,v_s) &=& 
    \frac{2}{\pi} \int_0^{\infty} \frac{dw}{w} \Delta \rho^{\rm shear}(w) .
    \label{correctedshearsumrule}
\end{eqnarray}
We will often employ the shorthand notation $\Delta$ to denote a
quantity which has the zero temperature part subtracted out.  
For example, 
\begin{equation}
  \Delta \rho(w) \equiv \rho(w) - \rho_{T=0}(w).
\end{equation}
To be consistent with currently known results, 
$F(\varepsilon, P, v_s)$ must vanish when $\varepsilon = 3P$ and 
$v_s^2 = 1/3$.

Using the same techniques as in \cite{Romatschke:2009ng}, we have been
able to derive the correction of the left hand side of the sum rule in
our particular non-conformal gravity dual.  We explicitly show that the
left and right sides of our corrected sum rule
(\ref{correctedshearsumrule}) agree within the numerical error.

The non-conformal theory in which we work is a simple 5D single scalar
model with an exponential potential.  This model is sometimes called
the Chamblin-Reall model \cite{Chamblin:1999ya}, and has been
extensively studied in the literature \cite{Springer:2008js,
  Springer:2009wj, Gubser:2008ny, Gubser:2008yx, Kanitscheider:2008kd,
  Romatschke:2009kr, Bigazzi:2010ku}.  We emphasize that this model is
not particularly well suited for QGP phenomenology; it has no
conserved charge, and also has the peculiar feature of being both
non-conformal and having a speed of sound which is independent of
temperature.  Still, we choose to work in this model because it is
perhaps the simplest example of a non-conformal gravity dual where
many of the hydrodynamic equations can be solved exactly.  It is worth
mentioning that \emph{if} there is a precise field theory dual to this
model, it is not known at present.  However, recently it was found
that the dynamics of a more complicated string theory setup (including
fundamental flavors) were captured by an effective single scalar
Chamblin-Reall background \cite{Bigazzi:2009tc}.  This may indicate a
connection between the Chamblin-Reall background and more rigorous
non-conformal deformations of $\mathcal{N} = 4$ SYM theory.
\commentout{It is also
worth mentioning that recently it was found that the dynamics of a
more complicated string theory setup (including fundamental flavors)
were captured by an effective single scalar Chamblin-Reall background
\cite{Bigazzi:2009tc}.  This may be an indication that there is a
stronger connection between the Chamblin-Reall background and string
theory than one might expect.}

Our paper is organized as follows.  In Sec. \ref{background_section},
we introduce our non-conformal gravitational dual, the single scalar
Chamblin-Reall background.  In Sec. \ref{RHS_section}, we present the
details of the evaluation of the right hand side of the sum rule.
This section involves introducing a tensor perturbation into the
geometry, numerically solving for the spectral density, and
integrating the result.  In Sec. \ref{LHS_section}, we evaluate the
left side of the sum rule using the known thermodynamics of the
gravity background; it is evident that the left side does not agree
with the right side except in the limiting case of a conformal theory.
We then proceed to derive the correct form of the left side and
present an improved sum rule where the left and right sides agree
numerically.  We conclude the paper in Sec. \ref{Conclusion_section}.
In Appendix \ref{app_YM} we discuss the relevant sum rule within the
context of (weakly coupled) Yang-Mills theory.  Other technical
details of our calculations and useful reference formulae are found in
Appendices \ref{app_BGequations} - \ref{Hinf_app}.

\section{Gravity background}
\label{background_section}
The theory under consideration is a 5D gravitational dual generated by
a single scalar field\footnote{Throughout this work, we use the
``mostly plus'' metric signature.}
\begin{eqnarray}
	\mathcal{S} &=& \frac{1}{2 \kappa} \int \, d^5 x \sqrt{-g} 
	\left[
	  R - \frac{1}{2} \partial_\mu \phi \partial^\mu \phi - V(\phi) 
	\right] \nonumber \\ 
	&+& \frac{1}{\kappa} \int d^4 x \sqrt{-\gamma} \theta,
\label{action}
\end{eqnarray}
where $\kappa$ is related to the five dimensional Newton's constant, 
$\kappa \equiv 8 \pi G_5$.
The second term is a boundary contribution, the well known
Gibbons-Hawking term which is necessary for a well defined variational
principle.  The induced metric on the boundary is denoted by
$\gamma_{\mu \nu}$, $\nabla_\mu$ denotes the covariant derivative, and
$\theta$ is the trace of the second fundamental form
\be
	\theta_{\mu \nu} = \nabla_\mu \hat{n}_{\nu}
\ee 
with $\hat{n}^{\nu}$ a unit vector normal to the boundary.  

We will assume the metric is of the ``black brane'' type
\be
ds^2 = g_{tt}(z)dt^2 + g_{xx}(z) d\mathbf{x}^2 + g_{zz}(z) dz^2,
\ee
and that the coordinates can be chosen such that there is a black brane horizon at
$z = z_h$.  We will often employ the symbol
\be
f(z) = -g_{tt}(z) g^{xx}(z).
\label{fdef}
\ee
As mentioned in the introduction, in presenting our main results we
will specify to a particular type of metric, the Chamblin-Reall
background.  However, whenever possible, we will keep the metric
components general in hopes that doing so may be useful for those
wishing to do analogous calculations in different backgrounds.

The Chamblin-Reall background can be found by assuming an exponential potential of the form
\begin{eqnarray}
  V(\phi) = -\frac{6}{L^2} \frac{(2-\delta)}{(1-2\delta)^2} \exp \left\{{\sqrt{\frac{4 \delta}{3}}} \phi \right\}.
\end{eqnarray}
The potential contains a parameter $\delta$ related to conformal symmetry breaking; the precise form
of the potential above is chosen for future convenience.  
The resulting metric and scalar field profile which solve Einstein's equations are
\begin{eqnarray}
  ds^2 &=& b^2(z) \left[ -f(z) dt^2 + d\mathbf{x}^2 + \frac{dz^2}{f(z)} \right] \label{zmetric}\\
  b(z) &=& \left(\frac{L}{z}\right)^{\frac{1}{1-2 \delta}}\\
  f(z) &=& 1-\left(\frac{z}{z_h}\right)^{\frac{2(2-\delta)}{1-2\delta}}\\
  \phi (z) &=& -\sqrt{12 \delta} \,\log [b(z)].
\end{eqnarray}
Here $L$ is a constant which is related to the radius of curvature of the space, and $z_h$ is the 
position of the horizon.  The coordinate $z$ runs from 0 to $z_h$, with the UV boundary at $z = 0$.  

Thermodynamics and transport coefficients have been studied in this setup in 
\cite{Springer:2008js,Springer:2009wj,Kanitscheider:2009as, Romatschke:2009kr, Bigazzi:2010ku}.  The relevant
results for our purposes are
\begin{eqnarray}
  \varepsilon &=& \frac{3}{1-2\delta} P = \frac{3}{2(2-\delta)} T s  \label{energydensity}\\
  v_s^2 &=& \frac{1}{3}(1-2\delta) \\
  \frac{\zeta}{\eta} &=& 2\left(\frac{1}{3} - v_s^2\right)
\end{eqnarray}
Here we have introduced $s$ as the entropy density, $P$ as the pressure, 
and $v_s$ as the speed of sound.  Note that the
parameter $\delta$ is a measure of the conformal symmetry breaking; for
$\delta = 0$, we recover the usual $AdS_5$ metric, which is dual to a
conformal field theory.  One should also note that this setup is
rather peculiar in that the speed of sound is \emph{constant} with
respect to temperature, though it is not necessarily equal to $1/\sqrt{3}$.
We will always work in the regime where $0 \leq \delta < 1/2$; in this
regime the speed of sound is positive and less than $1/\sqrt{3}$.

\section{Right side of sum rule}
\label{RHS_section}

\subsection{Tensor mode perturbations}
In order to access the two point correlation functions, we must add
perturbations to this geometry.  We assume a perturbation which depends
only on time and the extra-dimensional coordinate $z$.
\begin{eqnarray}
  g_{\mu \nu} &\rightarrow& g_{\mu \nu} + h_{\mu \nu}(t,z) \\
  \phi &\rightarrow& \phi_0 + \delta\phi(t,z).
\end{eqnarray}
In general, we might also assume a spatial dependence for the
perturbations.  Upon Fourier transform, we would acquire a momentum
dependence of $e^{i \mathbf{k} \cdot \mathbf{x}}$.  The sum rule in
question involves the two point correlation functions at vanishing
spatial momentum so we have set $\mathbf{k}$ to zero; this is
equivalent to assuming the perturbation does not depend on the spatial
coordinates $x^i$.

Perturbations in the 4D fluid are generally categorized into scalar,
vector, and tensor modes denoting their transformation properties
under spatial rotations.  We will examine the tensor mode; this is the
mode which gives access to the shear viscosity.  In this case, the
only nonzero metric perturbation is $h_{xy}$, and we need not consider
the fluctuation $\delta \phi$, as it does not couple to the metric 
perturbation in this channel.

In order to compute the correlation functions one must solve the
linearized Einstein equations for the perturbation's profile.  Once
this is accomplished, one must plug the result back into the action
and use the prescription of Son and Starinets \cite{Son:2002sd} to get
the correlation functions.

The linearized Einstein equation of motion for the Fourier transform
\begin{equation}
  H(t,z) \equiv h^x_{y}(t,z) = \int \frac{dw}{2 \pi} H (w,z) e^{-i w t} 
\end{equation}
is,
\begin{equation}
  \frac{1}{\rootg g^{zz}} \partial_z \left[ \rootg g^{zz} H' \right] - w^2 g_{zz} g^{tt} H = 0. 
  \label{zEOM}
\end{equation}
Throughout this work, we use the prime to denote derivative with respect to the coordinate which
labels the extra dimension (in the case at hand, $z$).  
This equation needs to be solved with the ``incoming wave'' boundary condition which can be applied by 
making the ansatz
\begin{eqnarray}
  H(z) = f(z)^{-i \W /2} Y(z),
  \label{incomingwave}
\end{eqnarray}
and requiring that $Y$ is a regular function of $z$ at the horizon.
We have defined the customary dimensionless frequency 
\begin{equation}
\W \equiv w / (2 \pi T). 
\end{equation}
The solution for $H$ will contain one integration constant, which 
can be related to the boundary value of $H(z \to 0)$.  

Correlation functions of the operator dual to the fluctuation $H$ can be 
found from the on-shell action.
In order to access two point functions, one needs to expand the
gravitational action to second order in perturbation $H$.  Upon
application of the equations of motion (``on-shell''), the
action reduces to boundary terms, some of which arise due to integration by
parts.  We will not go through the details of this
procedure, but they can be found in complete generality in
\cite{Buchel:2004di,Gubser:2008sz, Kaminski:2009dh}.  Here, we present the
on-shell boundary terms for the theory in question: a gravity dual
with a black brane metric supported by a single scalar field.

In writing the following expressions, we have chosen to remove all
instances of the potential with the background equations of motion.  These equations
can be found in Appendix \ref{app_BGequations}.    
We have also made use of the fact that all
backgrounds generated by scalar fields only must satisfy
\cite{Springer:2009wj}
\be
	g_{zz}(z) = c_1 \left[g_{xx}(z)\right]^4 \frac{[f'(z)]^2}{f(z)}.
\ee
This constraint is a consequence of the background Einstein equations.  
Here, $c_1$ is a constant related to the temperature
\be
	c_1 = \frac{1}{(4 \pi T)^2 g_{xx}(z_h)^3}.
\ee
Combining these expressions with the Beckenstein-Hawking entropy law
$s = g_{xx}(z_h)^{3/2} / 4 G$ allows one to remove Newton's
gravitational constant $G$ in favor of the thermodynamic quantities
$T, s$, and the constant $c_1$.  In computing the on-shell action, we
find the constant $c_1$ drops out completely.  The on-shell action is
divergent and so we introduce $\epsilon$ as a UV cutoff\footnote{One should take care
to distinguish the UV cutoff $\epsilon$ from the energy density $\varepsilon$.}.  There are two 
contributions to the on-shell action which can be written in terms of the quantity
\begin{widetext}
\begin{equation}
  \mathcal{S}\Bigl[A(z),B(z)\Bigr] \equiv \frac{T s V}{2} \int_{z = \epsilon}
  \frac{dw}{2 \pi} \frac{f(z)}{f'(z)} \Bigl\{ A(z) H'(w,z)H(-w,z) + B(z) H(w,z) H(-w,z) \Bigr\}.
\end{equation}
\end{widetext}
We have introduced $V$ to indicate the spatial volume, the result of
the integration over $d^3x$.
One contribution to the on-shell action is the results from the fact that bulk action (the first
term in (\ref{action})) reduces to a total derivative upon application of the equations 
of motion:
\begin{eqnarray}
  \mathcal{S}_{\rm bulk} = \mathcal{S}\Bigl[-3,-\DL[g_{xx}(z)] \Bigr].
\end{eqnarray}
We often employ the notation $\DL$ to denote the logarithmic derivative
\begin{eqnarray}
  \DL[X] = X'/X.  
  \label{DLdef}
\end{eqnarray}
In addition, there is a contribution from the Gibbons-Hawking term (the second term in (\ref{action})), which is
\begin{eqnarray}
  \mathcal{S}_{\rm GH} = 
  \mathcal{S} \Bigl[4,\DL[f(z)g_{xx}^4(z)] \Bigr].
\end{eqnarray} 
In total, then
\begin{equation}
  \mathcal{S}_{\rm Total} = \mathcal{S}_{\rm bulk} + \mathcal{S}_{\rm GH} = \mathcal{S}\Bigl[1,\DL[f(z)g_{xx}^3(z)] \Bigr].
\label{TotalAction}
\end{equation}
In general, the on-shell action needs to be regularized with the addition of counter terms.  However, 
if one is only interested in the imaginary part of the correlators (the spectral density $\rho$), this
is not necessary.  It is well known by now that the imaginary part of the on-shell action 
independent of the radial coordinate and is not divergent.  The imaginary part is\footnote{
This is because the on-shell action $\mathcal{S}$ is related to the retarded Green's function $G_R$,
which satisfies $G_R(w)^* = G_R(-w)$.}
\begin{equation}
  \mbox{Im } \mathcal{S}_{\rm Total} = \frac{1}{2i} \left[\mathcal{S}_{\rm Total}(w) - \mathcal{S}_{\rm Total}(-w)\right] 
\end{equation}
which can be written
\begin{eqnarray}
  \mbox{Im } \mathcal{S}_{\rm Total} &=& \frac{T s V}{4i} \int
  \frac{dw}{2 \pi} \frac{f}{f'} \Bigl[ H'(w,z)H(-w,z) \Bigr. \nonumber \\
    &-& \Bigl.H'(-w,z)H(w,z) \Bigr].
\end{eqnarray}
As mentioned above, one can evaluate this quantity at any value of $z$; a convenient one is $z = z_h$.  We assume
that our function $f$ vanishes linearly at the horizon; 
\begin{equation}
f(z = z_h) = f_0 (z - z_h) + \mathcal{O}\bigl((z-z_h)^2\bigr), 
\end{equation}
with $f_0$ a constant.  Furthermore, 
note that the incoming wave boundary conditions require 
\begin{equation}
  H'(w,z_h)(z-z_h) = -\frac{i \W}{2}H(w,z_h) \left[ 1 + \mathcal{O}(z - z_h) \right]
\end{equation}
very near the horizon.  
Thus,
\begin{eqnarray}
  && \mbox{Im } \mathcal{S}_{\rm Total} 
  = -\frac{V T s}{4} \int  \frac{dw}{2 \pi} \W \times \nonumber \\
  &&H(-w, \epsilon) 
  \left[\frac{H(w,z_h)H(-w,z_h)}{H(-w, \epsilon)H(w, \epsilon)} \right] H(w,\epsilon) .
\end{eqnarray}
Here, $H(w, \epsilon)$ is the boundary value of the perturbation as $z \to 0$.  
The prescription of Son and Starinets states that the two point correlation function of the operator
dual to $H$ is given by \cite{Son:2002sd}
\begin{equation}
  -\mbox{Im } G_R^{\rm shear} = \frac{T s \W}{2} \left[\frac{H(w,z_h)H(-w,z_h)}{H(-w, \epsilon)H(w, \epsilon)} \right]
\end{equation}
or, equivalently,
\begin{eqnarray}
  \rho(w) = \frac{s w}{4 \pi} \left[\frac{Y(w,z_h)Y(-w,z_h)}{Y(-w, \epsilon)Y(w, \epsilon)} \right] \label{rhoformula}.
\end{eqnarray}

In the case of $w = 0$, the only solution to the equations of motion which obeys the boundary conditions is
 $H = \rm{constant}$, and thus,
\begin{eqnarray}
  \eta = \lim_{w \to 0} \frac{\rho(w)}{w} = \frac{s}{4 \pi}  
\end{eqnarray}
The first equality is the Kubo formula for the shear viscosity.  This
is now a familiar result.  

\subsection{Numerical computation of spectral density}
We are interested in the quantity $\rho$ at
finite values of $w$, and thus the equations of motion must be solved
numerically.  The first step towards this end is to pass to a more convenient coordinate system.  
We define a dimensionless coordinate $u$ such that the horizon is at $u = 1$ and the boundary
is at $u = 0$.  One can write the metric as:
\begin{eqnarray}
  ds^2 &=& \frac{u^{\frac{2}{\delta-2}}}{\alpha} \left[ -dt^2 f(u) + d\mathbf{x}^2 \right] + \frac{u^{\frac{2(2+\delta)}{\delta-2}}}{\alpha (2 \pi T)^2} \frac{du^2}{f(u)}\\
f(u) &=& 1-u^2
\end{eqnarray}
The constant $\alpha$ is not important for our purposes (it drops out of the equations of motion), but for completeness, it is
\begin{eqnarray}
	\alpha = \left[ \frac{2(2-\delta)}{4 \pi T L(1-2 \delta)} \right]^{\frac{2}{1-2\delta}}.    
\end{eqnarray}
In this coordinate system, the equation of motion (\ref{zEOM}) becomes
\begin{equation}
  H''(u) - \frac{1+u^2}{u(1-u^2)}H'(u) + \W^2 \frac{u^{\frac{2(1+\delta)}{\delta-2}}}{(1-u^2)^2} H(u) = 0.
\end{equation}
Upon insertion of the incoming wave ansatz (\ref{incomingwave}), the equation for $Y$ is
\begin{equation}
  Y''-\frac{1+u^2 (1-2 i \W)}{u \left(1-u^2\right)}Y'-\frac{\left(1-u^{\frac{6}{\delta-2 }}\right) u^2 \W^2}{\left(1-u^2\right)^2}Y=0.
\label{YEOM}
\end{equation}

We use Mathematica's NDSolve function \cite{Mathematica} to solve this equation
numerically for a given value of $\W$ and $\delta$.  Boundary
conditions must be specified.  The function $Y$ must be regular at the
horizon in order to comply with the incoming wave boundary condition;
the easiest way to apply this condition is to begin integration at
some value of $u$ close to the horizon and specify
\begin{equation}
Y(1-\epsilon) = 1.
\label{BC1}  
\end{equation}
One also needs to specify the derivative $Y'(u)$ here.  Expanding the
equation (\ref{YEOM}) in powers of $(1-u)$, one finds that the leading order
term leads to the condition
\begin{eqnarray}
  Y'(1-\epsilon) = \frac{3 i \W^2 }{2 (i+\W) (2-\delta )}Y(1-\epsilon).  
\label{BC2}
\end{eqnarray}
To summarize, our numerical method is as follows
\begin{enumerate}
  \item 
    Specify a value of $\delta$ and $\W$.
  \item
    For these values of $\delta, \W$, use Mathematica's NDSolve to numerically integrate
    (\ref{YEOM}) with the boundary conditions (\ref{BC1}),(\ref{BC2}).  Start the integration near the
    horizon at $u = 1-\epsilon$, and integrate down very near the boundary at $u = \epsilon$.  
  \item
    Using the now known values of $Y(\epsilon)$ and (\ref{rhoformula}) one can determine the
    spectral density\footnote{One may be worried about applying our results to this new coordinate system; in fact, 
      the only coordinate dependent assumption we have made is that the $g_{tt}$ component vanish linearly 
      near the horizon, and the $g_{zz}$ or $g_{uu}$ component diverge as $1/x$ near the horizon.  These facts
      are true in both coordinate systems.}.  
    \begin{eqnarray}
      \frac{\rho^{\rm shear}(\mathfrak{w})}{\mathfrak{w}} 
      &=& \frac{T s}{2} \frac{1}{\left|Y(\epsilon)\right|^2}.
      \label{rhoformulaU}
\end{eqnarray}
\end{enumerate}

\subsection{Zero temperature subtraction}
The quantity that enters the sum rule is the zero temperature subtracted 
spectral density.  One can compute $\rho^{\rm shear}(w)_{\rm T=0}$
analytically.  A gravitational metric dual to a zero temperature field theory is, 
intuitively, one without a black brane horizon.  Returning now to our original
$z$ coordinates, we set $f(z) = 1$.  In terms of $b(z)^2 = g_{xx}(z)$, the equations
of motion become
\begin{eqnarray}
  H_0''(z) + \DL[g_{xx}^{3/2}(z)]H_0'(z) + w^2 H_0(z) = 0.
  \\ \nonumber 
\end{eqnarray}
We are using $H_0(z)$ to denote the solution at zero temperature.  We
are interested in the case where $g_{xx}(z) =
\left(\frac{L}{z}\right)^{n}$, with $n$ being a function of delta. 
\begin{eqnarray}
  n \equiv \frac{2}{1-2\delta}.
  \label{ndef}
\end{eqnarray}
 The equation is then
\begin{equation}
  H_0''(z) - \frac{3n}{2 z} H_0'(z) + w^2 H_0(z) = 0.
\label{bessleq}
\end{equation}
This equation is solved in terms of Bessel functions, or alternatively in terms of Hankel functions of the 
first and second kind:
\begin{eqnarray}
  H_0(z) = z^{\frac{2+3 n}{4}} \left[ C_1 h^{(1)}_{\frac{2+3n}{4}}(w z) + C_2 h^{(2)}_{\frac{2+3n}{4}}(w z) \right]. 
\label{H0soln}
\end{eqnarray}
The combination $(2+3n)/4$ appears frequently in our calculations, and for simplicity we will
use the definition
\begin{equation}
  l \equiv \frac{2+3n}{4} = \frac{2-\delta}{1-2\delta}.
  \label{ldef}
\end{equation}
The Hankel functions of the first kind behave at $z \to \infty$ as $\sim
e^{i w z} / \sqrt{z}$, whereas the Hankel functions of the second kind
behave as $\sim e^{-i w z} / \sqrt{z}$.  One can think of the zero
temperature metric as possessing a ``horizon'' at $z = \infty$, hence we
should choose $C_2 = 0$ so that waves are only traveling towards the
``horizon'' \cite{Son:2002sd}.

To get the correlation functions, we again need to expand the zero
temperature on-shell action to quadratic order in the perturbation
$H_0$.  The steps are analogous to those above, only now we are
working in a coordinate system where $-g_{tt} = g_{xx} = g_{zz}$.  The
results can be written in terms of the quantity
\begin{widetext}
\begin{equation}
    \mathcal{S}^{\rm T=0}\Bigl[A(z),B(z) \Bigr] = \frac{s V}{8 \pi} \int_{z=\epsilon} \frac{dw}{2 \pi} 
    \left(\frac{g_{xx}(z)}{g_{xx}(z_h)} \right)^{3/2} 
    \left[ A(z) H_0'(w,z)H_0(-w,z) + B(z) H_0(w,z)H_0(-w,z) \right]. 
\end{equation}
\end{widetext}
The results for the bulk and Gibbons-Hawking terms are
\begin{eqnarray}
  \mathcal{S}_{\rm bulk}^{\rm T=0} &=& 
  \mathcal{S}^{\rm T=0}\Bigl[3, \DL[g_{xx}(z)] \Bigr], \\
  \mathcal{S}_{\rm GH}^{\rm T=0} &=& 
  \mathcal{S}^{\rm T=0}\Bigl[-4, -\DL[g_{xx}^4 (z)] \Bigr], \\
  \mathcal{S}_{\rm Total}^{\rm T=0} &=& 
  -\mathcal{S}^{\rm T=0}\Bigl[1, \DL[g_{xx}^3 (z)] \Bigr].
\end{eqnarray}
And again applying the prescription of Son and Starinets, we find the spectral density,
\begin{align}
  &\rho^{\rm shear}_{\rm T=0}(w) = \frac{s}{8 \pi i} \left(\frac{g_{xx}(z_{*})}{g_{xx}(z_h)} \right)^{3/2} \times \nonumber  \\
  &  \left[ \frac{H_0'(w,z_{*})H_0(-w,z_*) -  H_0'(-w,z_*)H_0(w,z_*)}{H_0(w,\epsilon) H_0(-w,\epsilon)}  \right], \end{align}
The symbol $z_*$ is used to denote any particular value of $z$ which we choose (again, this result is independent of $z$).
In computing the finite temperature spectral density, we found it most convenient to evaluate the result at the horizon; here
it is more convenient to choose $z_* = \epsilon$.  Let us also now specify to the case at hand with $g_{xx} = (L/z)^n$.  
\begin{eqnarray}
  &\,& \rho^{\rm shear}_{\rm T=0}(w) = \frac{s}{8 \pi i} \left(\frac{z_h}{\epsilon} \right)^{3n/2} \times \nonumber \\ 
  &\,&  \left[ \frac{H_0'(w,\epsilon)H_0(-w,\epsilon) -  H_0'(-w,\epsilon)H_0(w,\epsilon)}{H_0(w,\epsilon) H_0(-w,\epsilon)}  \right] ,
\end{eqnarray}
With the use of the solution (\ref{H0soln}), one finds
\begin{eqnarray}
   H_0'(w,\epsilon)H_0(-w,\epsilon) &-&  H_0'(-w,\epsilon)H_0(w,\epsilon) \\ 
   &=& -\frac{4 |C_1|^2 (-1)^{-3n/4}(\epsilon)^{3 n/2}}{ \pi }. \nonumber
\end{eqnarray}
Finally, one needs to employ the expansion 
\begin{eqnarray}
  h^{(1)}_l(x) = -\frac{i}{\pi}\left(\frac{2}{x} \right)^l \Gamma(l) + \mathcal{O}(x^2) + \mathcal{O}(x^l).
\end{eqnarray}
Recall that $l >2$ in the physical region.   Putting it all together we find
\begin{eqnarray}
  \rho^{\rm shear}_{\rm T=0}(w) = \frac{ s w (w z_h)^{3 n/2}}{2^{2l+1}\Gamma(l)^2}.
\end{eqnarray}
Finally, one should remove $z_h$ in favor of $T$.  When the metric is written in the coordinate system (\ref{zmetric}), the 
Hawking temperature is
\begin{eqnarray}
  T = -\frac{f'(z_h)}{4 \pi} = \frac{l}{2 \pi z_h}.
\end{eqnarray}
Finally, using the definitions for $n$ and $l$ (\ref{ndef}) and (\ref{ldef}), we have our final result for the zero temperature case:
\begin{eqnarray}
  \frac{\rho^{\rm shear}_{\rm T=0}(\W)}{\W} 
  = \frac{ \pi T s \left[\frac{\W}{2} \left(\frac{2-\delta}{1-2\delta} \right)\right]^{\frac{3}{1-2\delta}}}
  {2 \Gamma\left(\frac{2-\delta}{1-2\delta}\right)^2}.
\end{eqnarray}
After numerically computing the spectral density at finite temperature, one subtracts off this piece to remove the 
UV divergences at large $\W$.  

One drawback of our method is that it requires a large degree of
numerical precision; both the zero temperature and finite temperature
spectral functions diverge at large $\W$, but their difference
approaches zero.  One needs to compute both spectral
functions to a high degree of precision before performing the
subtraction to get the desired result.  In the future it would be
desirable to build this subtraction into the numerics, to avoid this
need for very high precision numerics.

\subsection{Numerical results for spectral density}

\begin{figure}
\includegraphics[width=\columnwidth]{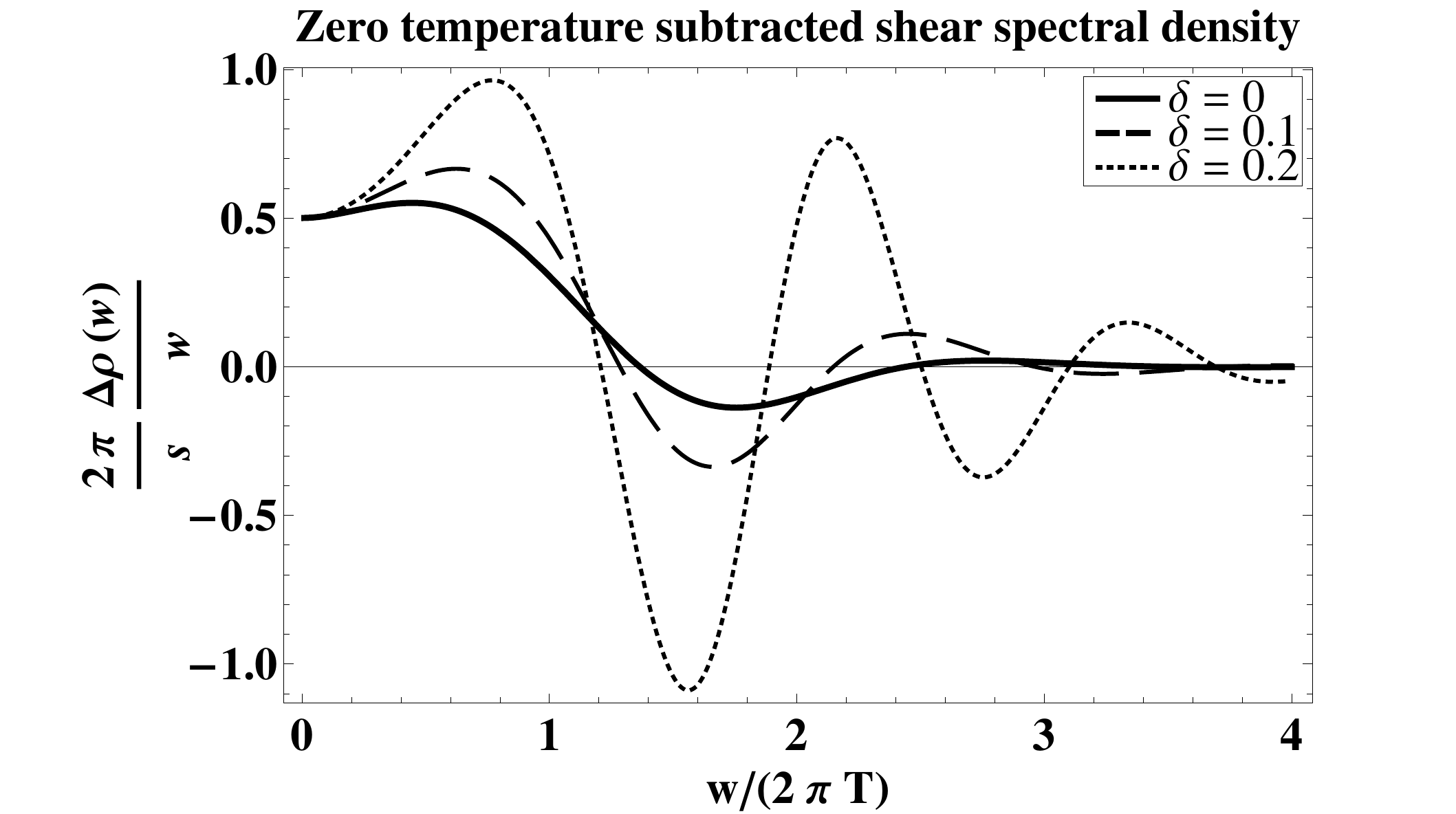}
\caption{
Plots of the zero-temperature subtracted spectral density versus
frequency for several values of $\delta$.  For large $w$, the
spectral density always approaches the zero temperature result.  For small $w$,
the slope of the spectral density always approaches the same result which is confirmation that
the shear viscosity takes on the universal value $\eta/s = 1/4\pi$.  
}
\label{spectralfigure}
\end{figure}

We have numerically computed the (zero-temperature subtracted) shear
spectral function for $\delta = 0, 0.1, 0.2$ and $0.3$.  Some sample
results are shown in Fig. \ref{spectralfigure}.  We have computed the
spectral functions out to a large value of $\W$; we cease our
numerical computation when the oscillations have reached 0.1\% of
their maximum value.  The qualitative behavior of this spectral
function is quite interesting.  The spectral density oscillates around
the zero temperature result, with the oscillations eventually dying
out as one moves to higher frequencies.  This oscillation phenomenon
has been noticed before in previous computations of the spectral
density in $\mathcal{N} = 4$ SYM theory
\cite{Kovtun:2006pf,Teaney:2006nc}.   These damped oscillations are
thought to be a reflection of the pole structure of the
retarded correlation functions in the complex plane \cite{Hartnoll:2005ju, Myers:2007we}.

\begin{figure}
\includegraphics[width=\columnwidth]{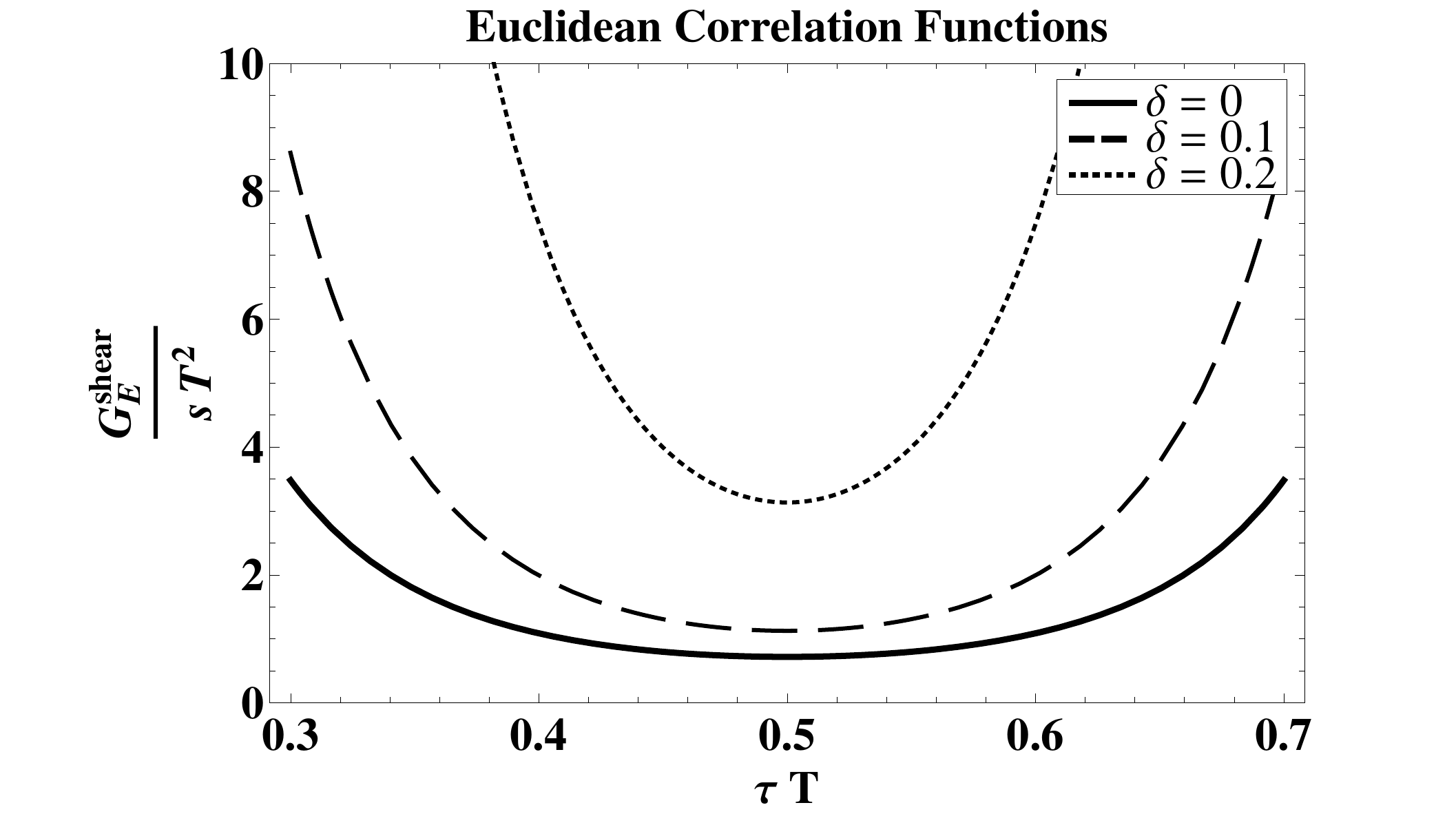}
\caption{ Plot of the Euclidean correlation function associated with
the shear spectral density as a function of the Euclidean time $\tau$
for various values of $\delta$.  The value of $\delta$ (and hence the
value of the bulk viscosity in our model) has a strong effect on the
shape and magnitude of these functions.  }
\label{GEfig}
\end{figure}

When moving to a non-conformal theory, we notice a qualitatively new
behavior.  As one increases the non-conformal parameter $\delta$, the
oscillations become initially larger and larger, and the spectral
function takes a longer time to settle down to its zero temperature
value.  This is especially evident for the case of $\delta = 0.3$
where the oscillations are so large that it does not fit nicely on a
plot with those shown in Fig. \ref{spectralfigure}.  We are unsure of
the physical interpretation of this behavior.
\begin{figure}
\includegraphics[width=\columnwidth]{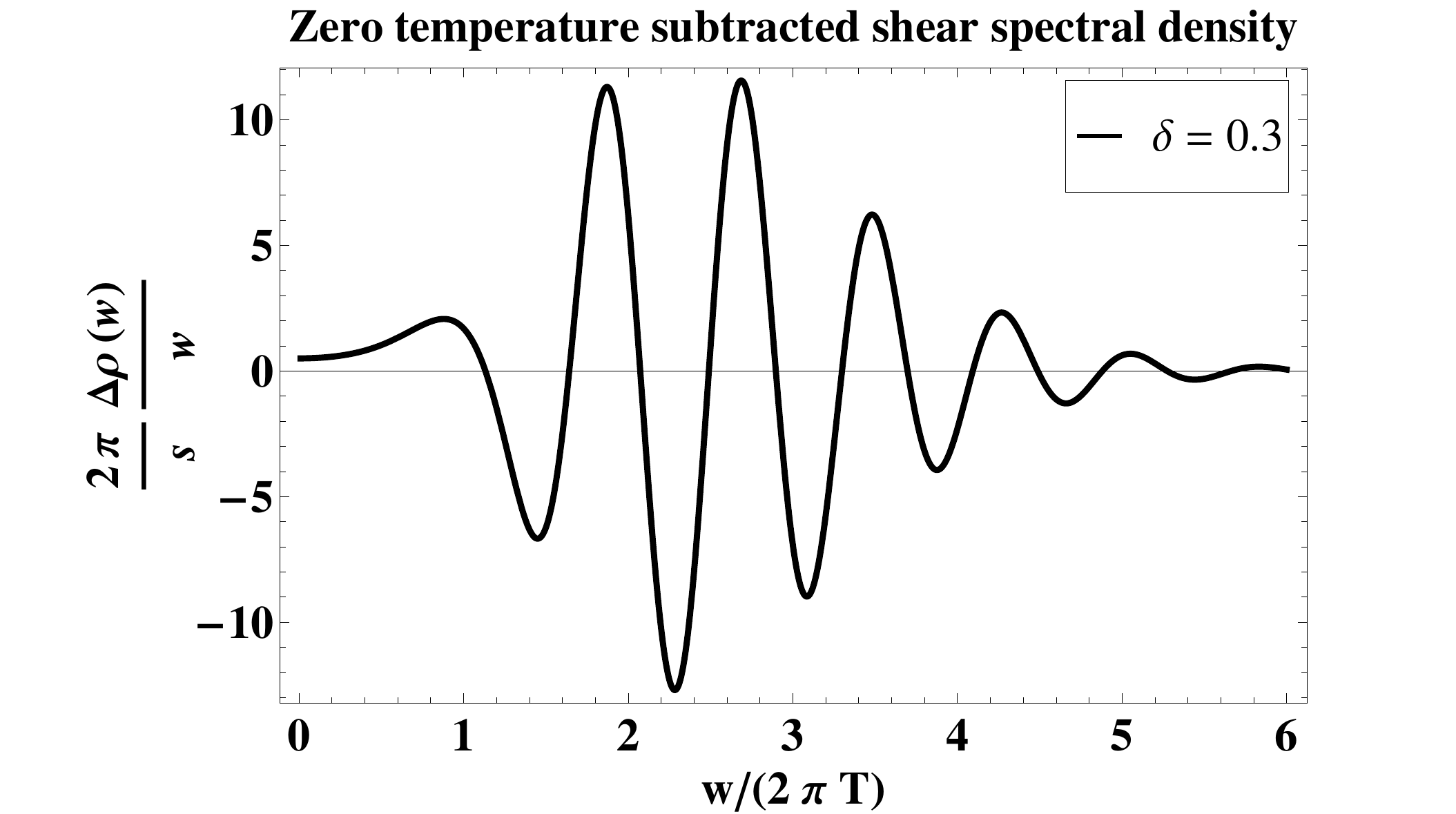}
\caption{
Plot of the zero-temperature subtracted spectral density versus
frequency for $\delta =0.3$.  One should note the qualitative
differences in the plots as one increases $\delta$.  For larger
values of $\delta$ the oscillations grow larger and more frequent, 
and take a longer time to die off.  Note the difference in the 
axis scaling between this plot and Fig. \ref{spectralfigure}.  
}
\label{spectralfigure2}
\end{figure}
However, we can use our spectral functions to compute Euclidean
correlation functions, quantities that are computed on the lattice (see for example \cite{Nakamura:2004sy, Meyer:2007ic}).
The relation between the spectral function and the Euclidean
correlators is
\begin{equation}
  G_E(\tau) = \frac{1}{\pi} \int \, dw \rho(w) \frac{\cosh\left[w(\tau - \beta/2) \right]}{\sinh \left[w \beta/2 \right]},
\end{equation}
where $\tau$ is the Euclidean time variable, which has period $\beta
\equiv 1/T$.  Using our numerical results for the spectral density, we
plot the results for the Euclidean correlation functions in
Fig. \ref{GEfig}.  It is interesting to note that the value of
$\delta$ (which is proportional to the bulk viscosity $\zeta$) affects
the magnitude of the Euclidean correlation function quite strongly.
This may have implications for lattice measurements, since the
Euclidean correlation function is directly measured there.  The
gravity dual in which we are working is only toy model of a
non-conformal theory, and hence we will not make any attempt to
extract a value of the bulk viscosity from the lattice data.  However,
\emph{if} the qualitative behavior noted here persists in more
realistic holographic models of physical gauge theories, it suggests
that one could possibly gain insight into the value of the bulk
viscosity from the \emph{tensor} correlation function considered here.
This could prove to be quite useful, since it would provide an
independent measurement of the bulk viscosity which is usually
extracted via the Kubo relations in the \emph{bulk} channel.

To evaluate the sum rule, one must integrate the spectral functions.
Technically, we use Mathematica to perform a cubic interpolation
between the points which are spaced at intervals of $\W = 0.01$, and
numerically integrate the resulting function.  The results are given
in Table \ref{SonTable}.

\section{Left side of sum rule}
\label{LHS_section}
If we take the sum rule (\ref{shearsumrule}) at face value, we can easily evaluate the left hand side 
using (\ref{energydensity}).  The results are shown in Table \ref{SonTable}.
\renewcommand{\arraystretch}{1.5}
\begin{table}
\begin{tabular}{|c|c|c|}
\hline
$\delta$ & (LHS of Sum rule)$\times \frac{1}{T s}$ & (RHS of Sum rule)$\times \frac{1}{T s}$ \\ 
\hline
0.0 & $\frac{3}{10} \approx 0.300...$ & $0.300...$  \\
\hline 
0.1 &  $\frac{6}{19} \approx 0.318...$ & $0.326...$   \\
\hline
0.2 &$\frac{1}{3} \approx 0.333...$ & $0.357...$  \\
\hline
0.3 & $\frac{6}{17} \approx 0.353...$& $0.395...$  \\
\hline
\end{tabular}
\caption{Results for the left and right sides of the sum 
rule given in \cite{Romatschke:2009ng}.  The left side values come from
(\ref{shearsumrule}) and (\ref{energydensity}).  The right side values come
from numerically integrating our results for the spectral density.  
\label{SonTable}
}
\end{table}
\renewcommand{\arraystretch}{1.5}
Clearly, there is substantial disagreement between the left and right
sides.  Notice that in the case of $\delta = 0$, the two sides are in
agreement, which was also the conclusion of \cite{Romatschke:2009ng}.
However once we deviate from conformality, differences appear.  The
error for $\delta = 0.3$ is greater than 10\% which is not accounted for
by numerical error.

Thus, we come to the conclusion that the left hand side of the sum
rule must be modified.  Following \cite{Romatschke:2009ng}, the sum
rule is, more generally
\begin{eqnarray}
  \Delta G_R^{\rm shear}(w = i \infty) &-& \Delta G_R^{\rm shear}(w = 0) \nonumber \\ 
  &=& \frac{2}{\pi} \int_0^{\infty} \frac{dw}{w} \Delta \rho^{\rm shear}(w).
\label{generalsumrule}
\end{eqnarray}
The left side of (\ref{generalsumrule}) can be computed using AdS/CFT
techniques.  The general method we use is the same as
\cite{Romatschke:2009ng}, except we are now working in a more general
background, and hence the equations of motion and action are modified.

We will first compute the term $\Delta G_R^{\rm shear}(w = i \infty)$.
Let us define
\be
	Q \equiv i w.
\ee
In terms of the dimensionless variable $\mathfrak{q} = i \W$, the
limit is $\mathfrak{q} \to \infty$.  Note that this limit can also be
achieved by taking $T \to 0$, or equivalently $z_h \to \infty$.  This,
in turn is equivalent to examining the metric in the regime near the
boundary $ z \to 0$.  Our first step is to transform the metric
to Fefferman-Graham like coordinates; the relevant details of this
transformation can be found in Appendix \ref{app_FG}.  Near the boundary,
the metric can be written as
\begin{eqnarray}
  ds^2 &=& \left(\frac{L}{\z}\right)^{\frac{2}{1-2\delta}} \left\{
  \left[-1 + \left(\frac{3-4 \delta }{4-4 \delta } \right)\left(\frac{\z}{z_h}\right)^{2l} \right]dt^2 \right. 
  \label{FGmetric}\\
  &+& \left[1+ \left(\frac{1}{4-4 \delta } \right)\left(\frac{\z}{z_h}\right)^{2l} \right]d\mathbf{x}^2 
  + \Biggl. d\z^2 \Biggr\} + \mathcal{O}\left(\frac{\z^{4l}}{z_h^{4l}} \right). \nonumber
\end{eqnarray}
One must now solve the equation of motion using this metric.  Since we
are working in the regime of $Q \to \infty$, we will denote the
relevant fluctuation as $H^{\infty}$.  Solving the equations of motion in this regime
is rather technical, but details can be found in Appendix \ref{Hinf_app}.  The
near boundary solution is
\begin{eqnarray}
\label{Hinfsoln}
  &\,& H^{\infty}(\z) \approx C_3 \left(\frac{\z}{z_h} \right)^l \times \\
  &\,& \left[ K_l(Q z) 
    -\left(\frac{\z}{z_h} \right)^l
    \left(\frac{2}{Q z_h}\right)^l 
    \frac{3 \Gamma(l)}{4(5-4\delta)}  + \mathcal{O}\left(\frac{\z^{2l}}{z_h^{2l}} \right)
    \right] \nonumber
\end{eqnarray} 
Here $C_3$ is a normalization constant, and $K_l$ is a Bessel function.  The reader is
referred to Appendix \ref{Hinf_app} for more details.  

To compute the correlator, we use the results for the on-shell action
given previously (\ref{TotalAction}), the Son and Starinets
prescription gives
\begin{equation}
  G_R(w = i\infty) = 
  T s \lim_{\epsilon \to 0}
  \frac{f(\epsilon)}{f'(\epsilon)} \frac{{H^{\infty}}'(Q,\epsilon)}{H^{\infty}(Q,\epsilon)}
  + G_R^{\rm CT}(w=i\infty).
\end{equation}
Here, the second term with the superscript $CT$ denotes the contact terms which arise from the part of the
on-shell action which contains no derivatives of $H$.  We will have
more to say about this term in a moment.  Our solution
(\ref{Hinfsoln}) for $H^{\infty}$ contains two terms; the first term
is the result if one takes the horizon $z_h \to \infty$; in other
words, it is the zero temperature result.  Upon subtracting the zero
temperature piece, we have
\begin{eqnarray}
\Delta G_R(w = i \infty) &=& T s \lim_{\epsilon \to 0}
  \frac{f(\epsilon)}{f'(\epsilon)} \frac{{H_1^{\infty}}'(Q,\epsilon)}{H_0^{\infty}(Q,\epsilon)}
  \nonumber \\
  &+& \Delta G_R^{\rm CT}(w=i\infty). 
\end{eqnarray}
Noting that
\begin{eqnarray}
  H^{\infty}_{1} = -\xi \left(\frac{\z}{z_h}\right)^{2l},
\end{eqnarray}
with $\xi$ a constant, and
\begin{eqnarray}
  f(\z) = 1 - \frac{\z^{2l}}{z_h^{2l}} + \mathcal{O}\left( \frac{\z^{4l}}{z_h^{4l}} \right)
\end{eqnarray}
we have,
\begin{equation}
  \Delta G_R(w = i \infty) = T s \frac{\xi}{H_0^{\infty}(Q,\epsilon)} 
  + \Delta G_R^{\rm CT}(w=i\infty).  
\end{equation}
Finally, with the fact that 
\begin{eqnarray}
  H^0_{\infty}(Q,\epsilon) = \frac{1}{2} \left(\frac{2}{Q z_h} \right)^l \Gamma(l) + \mathcal{O(\epsilon)}, 
\end{eqnarray}
we come to the result
\begin{equation}
  \Delta G_R(w = i \infty) = T s \left[\frac{3}{2(5-4\delta)} \right]
  + \Delta G_R^{\rm CT}(w=i\infty).  
\end{equation}
Ostensibly, we are not finished as the left side of the sum rule also
contains a term $\Delta G_R(w = 0)$, and we also have to deal with the
contact terms arising from the part of the action containing no
derivatives of $H$.  There is also the issue of counter terms which
could be added to regularize the action.  The claim of
\cite{Romatschke:2009ng} is that the contact terms above precisely
cancel the contribution at $w = 0$, and that all counter terms will
cancel if the zero temperature subtraction is done properly.  We will
not prove this claim here, but the agreement between our derived
formula and our numerical results is an empirical justification for
this claim.  We believe that showing this explicitly should be mostly
straightforward given our solutions and our expansions for the
on-shell action; however the zero temperature subtraction can
sometimes be a subtle issue (see for example, \cite{Gursoy:2008za}).
The final result of our analysis is 
\begin{equation}
  \Delta G_R(w = i \infty) - \Delta G_R(w = 0) = T s \left[ \frac{3}{2(5-4 \delta)} \right].  
\label{LHSfinal}
\end{equation}
Using now our improved formula we compare the left and right sides of
the sum rule in Table \ref{improvedtable}.
\renewcommand{\arraystretch}{1.5}
\begin{table}
\begin{tabular}{|c|c|c|}
\hline
$\delta$ & (\emph{New} Sum rule LHS)$\times \frac{1}{T s}$ & (Sum rule RHS)$\times \frac{1}{T s}$ \\ 
\hline
0.0 & $\frac{3}{10} \approx 0.300000$... & 0.3000(04)... \\
\hline 
0.1 & $\frac{15}{46} \approx 0.326087$...& 0.3260(86)... \\
\hline
0.2 & $\frac{5}{14} \approx 0.357143$...& 0.3571(28)...  \\
\hline
0.3 & $\frac{15}{38} \approx 0.394737$...& 0.395(036)... \\
\hline
\end{tabular}
\caption{\label{improvedtable} Comparison of the left and right sides
of the improved sum rule (\ref{sumruledelta}) which is now corrected to
include the non-conformal effects of the theory.  The left side is
computed from (\ref{LHSfinal}), while the right side is computed by the
numerical integration over the spectral density.  The parentheses
denote an estimate of numerical error which is explained more
completely in the text.}
\end{table}
\renewcommand{\arraystretch}{1}
As one can see, the agreement is much better, and the numerical results 
agree with the analytic ones to at least three significant figures.
The fact that the error increases with $\delta$ is not
surprising, since the numerics become more challenging as $\delta$
increases.  This is because the power divergences of $\rho$ and
$\rho_{\rm T = 0}$ are worse, and hence greater numerical precision is
required.  Furthermore, as $\delta$ increases, the oscillations take
longer to die out, and hence one must compute $\Delta \rho$ to a
larger value of $\W$ in order to achieve the same accuracy.  

The largest source of error in these computations is due to the fact
that we only compute $\rho$ up to a value $\W_{\rm max}$, and our
numerical integration stops at this value.  As shown in
\cite{Romatschke:2009ng}, the zero temperature subtracted spectral
density is an oscillating function which dies out nearly exponentially
at large $\W$.  We are neglecting the integral of this function from
$\W = \W_{\rm max}$ to $\W = \infty$.  One can estimate the
contribution of this ``tail'' by fitting the last few oscillations
near $\W_{\rm max}$ to a damped sine curve\footnote{This functional
  form was chosen for the purpose of error estimation only, and
  appears to fit the numerics quite well.  In fact, the form we have
  chosen matches an analytical calculation of a current-current
  correlation function in \cite{Myers:2007we}.  However, we stress
  that the exact functional form of $\Delta \rho^{\rm shear}$ is not
  known analytically at large $\W$, though it should be possible to
  calculate it using methods similar to those given in this paper. }
of the form
\begin{equation}
  \frac{\Delta \rho^{\rm shear}(\W \gg 1)}{\W} \approx a e^{-b \W} \sin \left( c \W + d \right).  
\end{equation}
Once the parameters $a,b,c$ and $d$ are found from the fit, one can then
integrate this function from $\W_{\rm max}$ to $\infty$ to estimate its
contribution.  We find that the order of magnitude of this tail
contribution is $\sim 10^{-5}$ for $\delta = 0, 0.1,$ and $0.2$.  The
contribution of the tail appears to increase with $\delta$, and is
roughly $\sim 10^{-4}$ for $\delta = 0.3$.  This is the significance of the
parenthesis in the table of our numerical results; it is expected that
the contribution of the tail will affect the numbers inside the
parenthesis.  However, it is quite clear that our analytic results
agree with the numerical ones within our estimated numerical error.  

The sum rule for the particular non-conformal theory
in which we are working is
\begin{equation}
	 T s \left[ \frac{3}{2(5-4 \delta)} \right] 
	 = \frac{2}{\pi} \int_0^{\infty} \frac{dw}{w} \Delta \rho^{\rm shear}(w).  
\label{sumruledelta}
\end{equation}
It is desirable to write the left side in terms of thermodynamic
observables, in hopes that our sum rule could be applicable to other
theories beyond those considered here.  Unfortunately, there is not a
unique way of doing that in our case, as one could write either
\begin{equation}
  \delta = \frac{\varepsilon - 3P}{2 \varepsilon}\, \mbox{   or    }\,   \delta = \frac{1}{2}\left(1-3 v_s^2 \right).
\end{equation}
This is a consequence of the fact that the speed of sound is
independent of temperature in this model.  One could perhaps gain more
information by investigating this sum rule in a gravitational
dual theory where the speed of sound depends on temperature
(e.g. \cite{Gursoy:2008bu,Gursoy:2008za,Gubser:2008yx,Gubser:2008ny})

We elect to write the left side in terms of $\varepsilon$ and $P$ only.  
Then, the sum rule can be written
\begin{equation}
  \frac{\varepsilon  (\varepsilon+P )}{2\left(\varepsilon+2P
  \right)}
  =\frac{2}{\pi} \int_0^{\infty} \frac{dw}{w} \Delta \rho^{\rm shear}(w). 
\label{Finalsumrule} 
\end{equation}
This sum rule reduces to (\ref{shearsumrule}) in the case of $\varepsilon = 3P$.
In other words, we have found that the function $F(\varepsilon, P, v_s)$ defined in (\ref{correctedshearsumrule})
can be written
\begin{equation}
  F(\varepsilon,P,v_s) = \frac{\varepsilon (\varepsilon - 3P)}{10(\varepsilon + 2P)}.
\end{equation}
\section{Conclusion}
\label{Conclusion_section}
In this work, we have examined a sum rule involving a particular two
point function of the energy-momentum tensor.  A version of this sum
rule was derived in \cite{Romatschke:2009ng}, but we have explicitly
shown that the sum rule given there is not valid for all Einstein
gravity dual theories.  We have provided a non-conformal generalization
of this sum rule; the main result of this work is
(\ref{Finalsumrule}).  We have numerically verified that the left side
of our improved sum rule equals the right side with an accuracy
greater than 0.1\%.  Whether our result is applicable to other
non-conformal gravity duals, or even to non-conformal field theories
should certainly be tested.  To this effect, we have examined the sum
rule in Yang-Mills theory in Appendix \ref{app_YM}.  In addition, it
is also important to numerically verify other sum rules given in
\cite{Romatschke:2009ng}.  These investigations are currently
underway.

Finally, in the course of our investigation of the sum rule, we
computed the spectral density in the tensor channel at various values
of our non-conformal deformation parameter $\delta$.  The behavior of
the spectral density and the associated Euclidean correlation
functions exhibit interesting qualitative behavior as a function of
$\delta$ as shown in Figs. \ref{spectralfigure} - \ref{GEfig}.  While
the qualitative features of the spectral density and Euclidean
correlation functions were found in the conformal case of
$\mathcal{N}=4$ SYM theory \cite{Kovtun:2006pf,Teaney:2006nc}, to the
best of our knowledge the change in these functions as a result of
non-conformality has not been examined before.  In particular, our
results seem to suggest that the bulk viscosity (which is proportional
to $\delta$) may have strong effects on the shape and magnitude of the
\emph{tensor} Euclidean correlators measured on the lattice.  This is
noteworthy because the bulk viscosity is usually found from the Kubo
relation which involves a different correlation function.  It remains
to be seen whether these qualitative features will persist in more
detailed gravitational dual models which attempt to capture more
features of the quark-gluon plasma.

\section*{Acknowledgments}
We thank Keshav Dasgupta, Mohammed Mia, Guy Moore, Paul Romatschke,
and Dam Son for helpful discussions.  This work was funded by the
Natural Sciences and Engineering Research Council of Canada.

\appendix

\section{Sum rule in Yang-Mills theory}
\label{app_YM}
We will use this section to make some comments on the
shear sum rule in (weakly coupled) Yang-Mills theory.  A tentative form
for this sum rule was written in \cite{Romatschke:2009ng}; but here we will
perform the calculation in a slightly different way.

The left side of the sum rule contains correlation functions at large frequency.  Such
quantities can be calculated using the operator product expansion (OPE).  In the limit of
$w \to \infty$, one only needs to know the OPE to leading order.  In \cite{CaronHuot:2009ns}, the leading order
OPE was calculated for the correlator of interest, with the result (in Euclidean signature)
\begin{equation}
  G^{\rm shear}(q \to \infty) = \frac{2}{3q^2} \left( q_4^2 - \vec{q}^{\,2} \right) \left< T^{44} \right>
  + \frac{1}{6} \left<F^2\right>
\label{Simonsformula}
\end{equation}
Here, $F_{\mu \nu}^a$ is the gluon field strength tensor, and $F^2
\equiv F_{\mu \nu}^a F^{\mu \nu}_a$.  The stress-energy tensor is denoted by $T^{\mu \nu}$,
\begin{equation}
  T_{\mu \nu} = F_{\mu \lambda}^{a} {F^a_{\nu}}^{\lambda} - \frac{1}{4}g_{\mu \nu} F^2,
\end{equation}
and the index ``4'' pertains to $q_4 = i w$.
In the limit of zero spatial momentum , and returning to Minkowski signature, we find
\begin{equation}
	G^{\rm shear}(w \to i\infty, \vec{q} \to 0) =  -\frac{2}{3} \left<T^{00}\right>
	+ \frac{1}{6}\left<F^2\right>.
\label{Simonsformulaq=0}
\end{equation}
All that remains is to express these quantities in terms of
thermodynamic ones; the arguments here follow \cite{Lee:2008xp} .
With the assumption of a perfect fluid, and with the use of the trace anomaly we can write
\begin{eqnarray}
  \left< T^{00} \right> &=& \varepsilon , \\
  \left< F^2 \right> &=& -\frac{2g}{\beta(g)} \left< T^\mu_\mu \right> =  \frac{2g}{\beta(g)}\left(\varepsilon - 3P \right), \label{F2}
\end{eqnarray}
where the scale dependence on the left hand side of (\ref{F2}) is transformed to the scale dependence of the coupling constant.
For notational convenience, we define
\begin{equation}
  G^{\rm shear}(w \to i\infty, \vec{q} \to 0) \equiv G^{\rm shear}_{\infty}.
\end{equation}
We find
\begin{equation}
  G^{\rm shear}_{\infty}
  = -\frac{2}{3} \varepsilon + \frac{g}{3 \beta(g)} \left(\varepsilon - 3P \right).
\end{equation}
To leading order, this becomes:
\begin{equation}
    -G^{\rm shear}_{\infty}
    = \frac{2}{3} \varepsilon
    + \frac{4 \pi}{11 N_c \alpha_s} \left(\varepsilon - 3P \right).
    \label{YMsumruleLHS}
\end{equation}
This is the quantity that will enter the left side of the sum rule for
pure gluodynamics, the low energy term $\Delta G_R(w = 0)$ vanishes as
shown in Appendix A of \cite{Romatschke:2009ng}.  Note that the second
term in (\ref{YMsumruleLHS}) which is proportional to $\alpha_s^{-1}$
was not given in \cite{Romatschke:2009ng}, though it appears to have
been independently noticed in a recent paper \cite{Meyer:2010gu}.
This additional term originates from the $\left<F^2\right>$ term in
\ref{Simonsformula}.  In \cite{CaronHuot:2009ns}, this term is argued
to be a contact term due to the fact that it renders the correlation
function non-transverse.  If this is the case, this additional term
should not be present in the sum rule, since all contact terms will
cancel out due to the subtraction $\Delta G_R(w = i \infty) - \Delta
G_R(w = 0)$.  Despite the claim made in \cite{CaronHuot:2009ns}, we
are unaware of any explicit calculation which shows that $\Delta G_R^{\rm shear}(w
= 0) \sim \varepsilon - 3P$.  For this reason, we currently choose to
include the additional term in the sum rule, though it is clear that
more work should be done to address this issue in the future.

In perturbation theory, $\varepsilon - 3P \sim
\mathcal{O}(\alpha_s^2)$, so in total, the second term on the right
side of (\ref{YMsumruleLHS}) is $\mathcal{O}(\alpha_s)$.  Let us make
a few comments on what may happen to this expression as we go beyond
leading order in perturbation theory.  We expect that the Wilson
coefficients in (\ref{Simonsformula}) will, in general, contain
corrections of $\mathcal{O}(\alpha_s)$.  In other words, we could
expect that (\ref{Simonsformula}) becomes
\begin{eqnarray}
  G^{\rm shear}(q \to \infty) &=& \frac{2}{3q^2}\left(1+a_1 \alpha_s\right) \left( q_4^2 - \vec{q}^{\,2} \right) \left< T^{44} \right> \nonumber \\
  &+& \frac{1}{6}\left(1+ a_2 \alpha_s \right) \left<F^2\right>
\label{Simonsformulanextorder}
\end{eqnarray}
with constants $a_1$ and $a_2$ to be determined from a one loop calculation.  Following through the
previous arguments, we would arrive at the expression
\begin{equation}
  -G^{\rm shear}_{\infty} = \frac{2}{3}\left(1+ a_1 \alpha_s \right)  \varepsilon
  + \frac{4 \pi}{11 N_c \alpha_s} \left(\varepsilon - 3P \right) + \mathcal{O}\left(\alpha_s^2 \right) .
  \label{YMsumruleLHSnextorder2}
\end{equation}
or, equivalently,
\begin{eqnarray}
  -G^{\rm shear}_{\infty} &=& \frac{1}{2}\left(1+ a_1 \alpha_s \right)\left( \varepsilon + P \right) \\
  &+& \frac{4 \pi}{11 N_c \alpha_s} \left(\varepsilon - 3P \right) + \mathcal{O}\left(\alpha_s^2 \right) \nonumber.
  \label{YMsumruleLHSnextorder}
\end{eqnarray}
where again we have used the fact that $\varepsilon - 3P$ is $\mathcal{O}\left(\alpha_s^2\right)$.

We find it interesting that the left side of the sum rule in our
strongly coupled model has a different dependence on $\varepsilon $
and $P$ than the left side of the sum rule in weakly coupled
Yang-Mills theory.  It is not clear whether this has any implications for the
possibility of a gravity dual of \emph{weakly coupled} Yang-Mills
theory.

\section{Background Equations}
\label{app_BGequations}
For the black brane type metric generated by a single scalar field, with
a $z$ denoting the radial coordinate, the background 
equations of motion can be written
\ba
\partial_z \left[ \rootg g^{zz} \DL[f] \right] &=& 0 
\label{BGEE1}\\
\frac{3}{2\sqrt{-g}} \partial_z \left[ \rootg g^{zz} \DL[g_{xx}] \right] &=& -\biggl. V(\phi) \biggr|_{\phi = \phi_0(z)} 
\label{BGEE2}\\
3 \DL[g_{xx}] \DL \left[ \frac{\sqrt{g_{zz} f}}{\DL[g_{xx}]} \right] &=& \phi_0'(z)^2 
\label{BGEE3}\\
\frac{1}{\rootg} \partial_z \left[ \rootg g^{zz} \phi_0'(z) \right] &=& \biggl. \frac{\partial V}{\partial \phi} \biggr|_{\phi = \phi_0(z)},
\label{BGEE4}
\ea
where $f$ and $\DL$ are defined as in the text (\ref{fdef}) and (\ref{DLdef}) respectively.

\section{Transformation to Fefferman-Graham like coordinates}
\label{app_FG}
In this section, we detail the transformation of the Chamblin-Reall metric to 
a Fefferman-Graham like coordinate system.  Fefferman-Graham coordinates are 
useful in many respects for asymptotically anti de-Sitter metrics; these
coordinates are defined
so that the metric takes the form
\begin{eqnarray}
  ds^2 = \frac{L^2 \left( \tilde{g}_{\mu \nu}dx^\mu dx^\nu + d\z^2 \right)}{\z^2},
\end{eqnarray}
where the indices $\mu$ and $\nu$ run over the four coordinates $t,x^i$.  

Note that the Chamblin-Reall metric
is not asymptotically anti de-Sitter (except for the case of $\delta = 0$),
hence we will make a slight generalization of the Fefferman-Graham coordinates,
where the metric is written
\begin{eqnarray}
  ds^2 = \frac{L^{n} \left( \tilde{g}_{\mu \nu}dx^\mu dx^\nu + d\z^2 \right)}{\z^{n}}.
\end{eqnarray}
It is not always possible to solve for $\tilde{g}_{\mu \nu}$
analytically, but one can find its behavior near the boundary as an
expansion in the radial coordinate.  To transform our metric to this
form, we begin with the Chamblin-Reall metric written in the
coordinate system \ref{zmetric}, and apply the coordinate
transformation \cite{Kajantie:2006hv},
\begin{eqnarray}
  \left(\frac{L}{z} \right)^n \frac{dz^2}{f(z)} = \left(\frac{L}{\z} \right)^n d\z^2. 
\end{eqnarray}
Enforcing the fact that near the boundary, $z = \z$, we can integrate
this equation to find
\begin{eqnarray}
  \int_\epsilon^z \frac{dz}{z^{n/2}\sqrt{f(z)}} =   \int_\epsilon^{\z} \frac{d\z}{\z^{n/2}}. 
\end{eqnarray}
For the time being, let us assume that $n \neq 2$, (though we will be
able to relax this assumption later on).  Then, performing the
integral we find
\begin{eqnarray}
  \z^{1-\frac{n}{2}} &=& \left(1-\frac{n}{2}\right) \int_\epsilon^z \frac{dz}{z^{n/2}\sqrt{f(z)}} + \epsilon^{1-\frac{n}{2}} 
  \\
  &=& z^{1-\frac{n}{2}} + \left(1-\frac{n}{2}\right) \int_0^z \frac{1}{z^{n/2}} \left[ \frac{1}{\sqrt{f(z)}} -1 \right] dz.  
  \nonumber \\
\end{eqnarray}
We have taken the limit $\epsilon \to 0$ freely since the integral
converges.  Near the boundary, one can expand the integrand to
find\footnote{We assumed $2l - \frac{n}{2} \neq -1$.  This is valid
provided $\delta \neq 5/2$ which is outside the physical regime of
interest}
\begin{eqnarray}
  \z^{1-\frac{n}{2}} &\approx& z^{1-\frac{n}{2}}\left\{1 + \left(1-\frac{n}{2}\right) \frac{1}{4l - n +2} \left(\frac{z}{z_h}\right)^{2l}
  \right.
  \nonumber \\
  &+& \left. \mathcal{O}\left(\frac{z^{2l}}{z_h^{2l}}\right) \right\}.   
\end{eqnarray}
(Note that $f(z) = 1- \left(z/z_h \right)^{2l}$.)  Then, to this order
we have,
\begin{eqnarray}
  \z &\approx& z \left\{1 + \frac{1}{4l - n +2} \left(\frac{z}{z_h}\right)^{2l} + \mathcal{O}\left(\frac{z^{4l}}{z_h^{4l}}\right) \right\}, \\
  z &\approx& \z \left\{1 - \frac{1}{4l - n +2} \left(\frac{\z}{z_h}\right)^{2l} + \mathcal{O}\left(\frac{\z^{4l}}{z_h^{4l}}\right) \right\}.
\end{eqnarray}
In terms of $\delta$, this is,
\begin{eqnarray}
  z &\approx& \z \left\{1 - \frac{1-2\delta}{8 - 8 \delta} \left(\frac{\z}{z_h}\right)^{2l} + \mathcal{O}\left(\frac{\z^{4l}}{z_h^{4l}}\right) 
  \right\}.
  \label{FGtransform}
\end{eqnarray}
Strictly speaking, we derived this result assuming $n\neq2$, or
equivalently $\delta \neq 0$.  However, if one redoes the analysis for
$\delta = 0$, one finds precisely the same result as that given by
(\ref{FGtransform}), thus (\ref{FGtransform}) is valid for all values
of $\delta$ in the physical regime $0 \leq \delta < \frac{1}{2}$.
Applying this coordinate transform, we find the Fefferman-Graham like
representation of the Chamblin-Reall metric
which is given in the text (\ref{FGmetric}) to order $\mathcal{O}(z^{2l}/z_h^{2l})$.
Since we are only interested in the near-boundary dynamics, we need
not worry about higher order terms.

\section{Calculation of $H^{\infty}$}
\label{Hinf_app}
In this section we present some of the technical details involved in
solving the equations of motion in the large $w$ regime.  Throughout
this section we work in the Fefferman-Graham coordinate system, which
is explained in detail in Appendix \ref{app_FG}.  
Perhaps the easiest way to get
the relevant equation is to introduce a scaling parameter $\lambda$,
by replacing $\left(\frac{\z}{z_h}\right)^{2l} \to \lambda
\left(\frac{\z}{z_h}\right)^{2l}$, making the ansatz
\begin{eqnarray}
  H^{\infty}(\z) = H^{\infty}_0(\z) + \lambda H^{\infty}_1(\z).  
\end{eqnarray}
and expanding the equation of motion (\ref{zEOM}) in $\lambda$.  The
lowest order equation gives the $T=0$ equation for $H_0$ (with $w^2$
replaced by $-Q^2$).
\begin{eqnarray}
   (H^{\infty}_0)''-\frac{3}{\z(1-2\delta )}(H^{\infty}_0)'-Q^2 H^{\infty}_0 = 0.
\end{eqnarray}
The first order equation is
\begin{eqnarray}
  (H^{\infty}_1)''&-&\frac{3}{\z(1-2\delta )}(H_1^{\infty})'-Q^2 H_1^{\infty} \\
  &=&\left(\frac{\z}{z_h}\right)^{2l} \frac{1}{1-\delta }\left[\frac{3-4 \delta }{4} Q^2 H^{\infty}_0
    - \delta l
    \frac{(H_0^{\infty})'}{\z}\right] \nonumber
\label{H1eqn}
\end{eqnarray}
The solution for $H_0$ is 
\begin{eqnarray}
  H_0^{\infty}(Q,\u) = C_3 \u^l K_l(Q \u z_h) + C_4 \u^l I_l(Q \u z_h), 
\end{eqnarray}
where we have defined $\u \equiv \z/z_h$ for convenience.  $I_l$ and $K_l$
are the modified Bessel functions of the first and second kind
respectively.  Regularity at $\z \to \infty$ \cite{Son:2002sd,
Romatschke:2009ng} requires $C_4 =0$.

The homogeneous part of the first order equation (\ref{H1eqn}) is the
same as the zero-temperature equation, hence this equation can be
solved with the use of a Green's function.  Defining the two
homogeneous solutions as
\begin{eqnarray}
  y_1(\u) &=& C_3 \u^l K_l(Q \u z_h) \\
  y_2(\u) &=& C_3 \u^l I_l(Q \u z_h),
\end{eqnarray}
The solution is
\begin{eqnarray}
  H_1^{\infty}(\u) &=& -\frac{1}{C_3^2} \left \{
  y_1(\u) \int_0^{\u} y_2(t) g(t)dt \right. \nonumber \\
  &+& \Biggl.  y_2(\u) \int_{\u}^{\infty} y_1(t) g(t)dt
  \Biggr\},
\end{eqnarray}
where 
\begin{eqnarray}
  g(\u) &=& \frac{\u}{1-\delta } \left[ 
    \frac{3-4 \delta }{4} (Q z_h)^2 H^{\infty}_0-\delta l \frac{(H^{\infty}_0)'}{\u}
    \right]\nonumber \\
  &=& \frac{\u}{1-\delta } \left[ 
    \frac{3-4 \delta }{4} (Q z_h)^2 y_1(\u)- \delta l\frac{y_1'(\u)}{\u}
    \right].
\end{eqnarray}
In order to compute the correlation function, we will only need the
leading term at the boundary $\u = 0$,
\begin{equation}
  H_1(\u \to 0) = -\frac{1}{C_3^2} \left \{
  y_2(u \to 0) \int_{0}^{\infty} y_1(t) g(t)dt
  \right\}
\end{equation}
which can be written:
\begin{eqnarray}
  H_1(\u \to 0) =
  \frac{1}{C_3 (\delta-1)} \left \{
  \left(\frac{\u^2 Q z_h}{2}\right)^l \frac{1}{l \Gamma(l)}\right. &\times&  \nonumber\\
  \Biggl. \int_{0}^{\infty} \left[ 
    \frac{3-4 \delta }{4} (Q z_h)^2 t y_1(t)^2-\frac{\delta l }{2}[y_1(t)^2]'
    \right]dt
  \Biggr\}. &&
\end{eqnarray}
The first term in the integrand requires the integral
\begin{eqnarray}
  \int_0^{\infty} dt K_l(t)^2 t^{2l + 1}  = \frac{ \Gamma(1+l) \Gamma(1+2 l)\sqrt{\pi }}{4 \Gamma\left(\frac{3}{2}+l\right)},
\end{eqnarray}
and the second term in the integrand is a total derivative which
reduces to the boundary term $\sim y_1(0)^2$, (note that $y_1(\u \to
\infty)$ approaches zero exponentially).  In total, the result is
\begin{eqnarray}
  && H_1^{\infty}(\z\to 0) = 
  \frac{C_3}{4(1-\delta)} 
  \left(\frac{\z^2}{2 Q z_h^3}\right)^l \nonumber \\
  &\times&
  \left[ 
    \frac{3 -4 \delta}{4}\frac{\Gamma(1+2 l)\sqrt{\pi }}{\Gamma\left(\frac{3}{2}+l\right)} 
    + \frac{\delta l 4^{l}}{2} \Gamma(l) 
    \right]. 
\end{eqnarray}
The identity
\begin{eqnarray}
  \Gamma(l+1) \Gamma\left(l + \frac{3}{2} \right) = \frac{\sqrt{\pi}}{2} \frac{1}{4^l} \Gamma(2l+2)
\end{eqnarray}
can be used to simplify the result to:
\begin{equation}
    H_1^{\infty}(\z\to 0) 
    = -C_3
    \left(\frac{2\z^2}{Q z_h^3}\right)^l \Gamma(l) \left(\frac{3}{4(5-4\delta)}\right).
\end{equation}
In summary, the solution for $H$ for large values of $Q$, $H(w =
i\infty) \equiv H_{\infty}$ can be written
\begin{eqnarray}
  &\,& H^{\infty}(\z) \approx C_3 \left(\frac{\z}{z_h} \right)^l \times \\
  &\,& \left[ K_l(Q z) 
    -\left(\frac{\z}{z_h} \right)^l
    \left(\frac{2}{Q z_h}\right)^l 
    \frac{3 \Gamma(l)}{4(5-4\delta)}  + \mathcal{O}\left(\frac{\z^{2l}}{z_h^{2l}} \right)
    \right] \nonumber
\end{eqnarray} 
This is the solution which is quoted in the text.

\bibliography{c:/Users/Todd/Documents/Physics/AdS_QCD/Drafts/Bibliography_Files/AdSCFT}

\begin{thebibliography}{42}%
\makeatletter
\providecommand \@ifxundefined [1]{%
 \@ifx{#1\undefined}
}%
\providecommand \@ifnum [1]{%
 \ifnum #1\expandafter \@firstoftwo
 \else \expandafter \@secondoftwo
 \fi
}%
\providecommand \@ifx [1]{%
 \ifx #1\expandafter \@firstoftwo
 \else \expandafter \@secondoftwo
 \fi
}%
\providecommand \natexlab [1]{#1}%
\providecommand \enquote  [1]{``#1''}%
\providecommand \bibnamefont  [1]{#1}%
\providecommand \bibfnamefont [1]{#1}%
\providecommand \citenamefont [1]{#1}%
\providecommand \href@noop [0]{\@secondoftwo}%
\providecommand \href [0]{\begingroup \@sanitize@url \@href}%
\providecommand \@href[1]{\@@startlink{#1}\@@href}%
\providecommand \@@href[1]{\endgroup#1\@@endlink}%
\providecommand \@sanitize@url [0]{\catcode `\\12\catcode `\$12\catcode
  `\&12\catcode `\#12\catcode `\^12\catcode `\_12\catcode `\%12\relax}%
\providecommand \@@startlink[1]{}%
\providecommand \@@endlink[0]{}%
\providecommand \url  [0]{\begingroup\@sanitize@url \@url }%
\providecommand \@url [1]{\endgroup\@href {#1}{\urlprefix }}%
\providecommand \urlprefix  [0]{URL }%
\providecommand \Eprint [0]{\href }%
\providecommand \doibase [0]{http://dx.doi.org/}%
\providecommand \selectlanguage [0]{\@gobble}%
\providecommand \bibinfo  [0]{\@secondoftwo}%
\providecommand \bibfield  [0]{\@secondoftwo}%
\providecommand \translation [1]{[#1]}%
\providecommand \BibitemOpen [0]{}%
\providecommand \bibitemStop [0]{}%
\providecommand \bibitemNoStop [0]{.\EOS\space}%
\providecommand \EOS [0]{\spacefactor3000\relax}%
\providecommand \BibitemShut  [1]{\csname bibitem#1\endcsname}%
\let\auto@bib@innerbib\@empty
\bibitem [{\citenamefont {Shifman}\ \emph {et~al.}(1979)\citenamefont
  {Shifman}, \citenamefont {Vainshtein},\ and\ \citenamefont
  {Zakharov}}]{Shifman1979385}%
  \BibitemOpen
  \bibfield  {author} {\bibinfo {author} {\bibfnamefont {M.~A.}\ \bibnamefont
  {Shifman}}, \bibinfo {author} {\bibfnamefont {A.~I.}\ \bibnamefont
  {Vainshtein}}, \ and\ \bibinfo {author} {\bibfnamefont {V.~I.}\ \bibnamefont
  {Zakharov}},\ }\href {\doibase DOI: 10.1016/0550-3213(79)90022-1} {\bibfield
  {journal} {\bibinfo  {journal} {Nucl. Phys.}\ }\textbf {\bibinfo {volume}
  {B147}},\ \bibinfo {pages} {385 } (\bibinfo {year} {1979})}\BibitemShut
  {NoStop}%
\bibitem [{\citenamefont {Adams}\ \emph {et~al.}(2005)\citenamefont {Adams}
  \emph {et~al.}}]{Adams:2005dq}%
  \BibitemOpen
  \bibfield  {author} {\bibinfo {author} {\bibfnamefont {J.}~\bibnamefont
  {Adams}} \emph {et~al.} (\bibinfo {collaboration} {STAR}),\ }\href {\doibase
  10.1016/j.nuclphysa.2005.03.085} {\bibfield  {journal} {\bibinfo  {journal}
  {Nucl. Phys.}\ }\textbf {\bibinfo {volume} {A757}},\ \bibinfo {pages} {102}
  (\bibinfo {year} {2005})},\ \Eprint {http://arxiv.org/abs/nucl-ex/0501009}
  {arXiv:nucl-ex/0501009} \BibitemShut {NoStop}%
\bibitem [{\citenamefont {Adcox}\ \emph {et~al.}(2005)\citenamefont {Adcox}
  \emph {et~al.}}]{Adcox:2004mh}%
  \BibitemOpen
  \bibfield  {author} {\bibinfo {author} {\bibfnamefont {K.}~\bibnamefont
  {Adcox}} \emph {et~al.} (\bibinfo {collaboration} {PHENIX}),\ }\href
  {\doibase 10.1016/j.nuclphysa.2005.03.086} {\bibfield  {journal} {\bibinfo
  {journal} {Nucl. Phys.}\ }\textbf {\bibinfo {volume} {A757}},\ \bibinfo
  {pages} {184} (\bibinfo {year} {2005})},\ \Eprint
  {http://arxiv.org/abs/nucl-ex/0410003} {arXiv:nucl-ex/0410003} \BibitemShut
  {NoStop}%
\bibitem [{\citenamefont {Back}\ \emph {et~al.}(2005)\citenamefont {Back} \emph
  {et~al.}}]{Back:2004je}%
  \BibitemOpen
  \bibfield  {author} {\bibinfo {author} {\bibfnamefont {B.~B.}\ \bibnamefont
  {Back}} \emph {et~al.} (\bibinfo {collaboration} {PHOBOS}),\ }\href {\doibase
  10.1016/j.nuclphysa.2005.03.084} {\bibfield  {journal} {\bibinfo  {journal}
  {Nucl. Phys.}\ }\textbf {\bibinfo {volume} {A757}},\ \bibinfo {pages} {28}
  (\bibinfo {year} {2005})},\ \Eprint {http://arxiv.org/abs/nucl-ex/0410022}
  {arXiv:nucl-ex/0410022} \BibitemShut {NoStop}%
\bibitem [{\citenamefont {Arsene}\ \emph {et~al.}(2005)\citenamefont {Arsene}
  \emph {et~al.}}]{Arsene:2004fa}%
  \BibitemOpen
  \bibfield  {author} {\bibinfo {author} {\bibfnamefont {I.}~\bibnamefont
  {Arsene}} \emph {et~al.} (\bibinfo {collaboration} {BRAHMS}),\ }\href
  {\doibase 10.1016/j.nuclphysa.2005.02.130} {\bibfield  {journal} {\bibinfo
  {journal} {Nucl. Phys.}\ }\textbf {\bibinfo {volume} {A757}},\ \bibinfo
  {pages} {1} (\bibinfo {year} {2005})},\ \Eprint
  {http://arxiv.org/abs/nucl-ex/0410020} {arXiv:nucl-ex/0410020} \BibitemShut
  {NoStop}%
\bibitem [{\citenamefont {Karsch}\ \emph {et~al.}(2008)\citenamefont {Karsch},
  \citenamefont {Kharzeev},\ and\ \citenamefont {Tuchin}}]{Karsch:2007jc}%
  \BibitemOpen
  \bibfield  {author} {\bibinfo {author} {\bibfnamefont {F.}~\bibnamefont
  {Karsch}}, \bibinfo {author} {\bibfnamefont {D.}~\bibnamefont {Kharzeev}}, \
  and\ \bibinfo {author} {\bibfnamefont {K.}~\bibnamefont {Tuchin}},\ }\href
  {\doibase 10.1016/j.physletb.2008.01.080} {\bibfield  {journal} {\bibinfo
  {journal} {Phys. Lett.}\ }\textbf {\bibinfo {volume} {B663}},\ \bibinfo
  {pages} {217} (\bibinfo {year} {2008})},\ \Eprint
  {http://arxiv.org/abs/0711.0914} {arXiv:0711.0914 [hep-ph]} \BibitemShut
  {NoStop}%
\bibitem [{\citenamefont {Kharzeev}\ and\ \citenamefont
  {Tuchin}(2008)}]{Kharzeev:2007wb}%
  \BibitemOpen
  \bibfield  {author} {\bibinfo {author} {\bibfnamefont {D.}~\bibnamefont
  {Kharzeev}}\ and\ \bibinfo {author} {\bibfnamefont {K.}~\bibnamefont
  {Tuchin}},\ }\href {\doibase 10.1088/1126-6708/2008/09/093} {\bibfield
  {journal} {\bibinfo  {journal} {JHEP}\ }\textbf {\bibinfo {volume} {09}},\
  \bibinfo {pages} {093} (\bibinfo {year} {2008})},\ \Eprint
  {http://arxiv.org/abs/0705.4280} {arXiv:0705.4280 [hep-ph]} \BibitemShut
  {NoStop}%
\bibitem [{\citenamefont {Ellis}\ \emph {et~al.}(1998)\citenamefont {Ellis},
  \citenamefont {Kapusta},\ and\ \citenamefont {Tang}}]{Ellis:1998kj}%
  \BibitemOpen
  \bibfield  {author} {\bibinfo {author} {\bibfnamefont {P.~J.}\ \bibnamefont
  {Ellis}}, \bibinfo {author} {\bibfnamefont {J.~I.}\ \bibnamefont {Kapusta}},
  \ and\ \bibinfo {author} {\bibfnamefont {H.-B.}\ \bibnamefont {Tang}},\
  }\href {\doibase 10.1016/S0370-2693(98)01292-1} {\bibfield  {journal}
  {\bibinfo  {journal} {Phys. Lett.}\ }\textbf {\bibinfo {volume} {B443}},\
  \bibinfo {pages} {63} (\bibinfo {year} {1998})},\ \Eprint
  {http://arxiv.org/abs/nucl-th/9807071} {arXiv:nucl-th/9807071} \BibitemShut
  {NoStop}%
\bibitem [{\citenamefont {Moore}\ and\ \citenamefont
  {Saremi}(2008)}]{Moore:2008ws}%
  \BibitemOpen
  \bibfield  {author} {\bibinfo {author} {\bibfnamefont {G.~D.}\ \bibnamefont
  {Moore}}\ and\ \bibinfo {author} {\bibfnamefont {O.}~\bibnamefont {Saremi}},\
  }\href {\doibase 10.1088/1126-6708/2008/09/015} {\bibfield  {journal}
  {\bibinfo  {journal} {JHEP}\ }\textbf {\bibinfo {volume} {09}},\ \bibinfo
  {pages} {015} (\bibinfo {year} {2008})},\ \Eprint
  {http://arxiv.org/abs/0805.4201} {arXiv:0805.4201 [hep-ph]} \BibitemShut
  {NoStop}%
\bibitem [{\citenamefont {Romatschke}\ and\ \citenamefont
  {Son}(2009)}]{Romatschke:2009ng}%
  \BibitemOpen
  \bibfield  {author} {\bibinfo {author} {\bibfnamefont {P.}~\bibnamefont
  {Romatschke}}\ and\ \bibinfo {author} {\bibfnamefont {D.~T.}\ \bibnamefont
  {Son}},\ }\href {\doibase 10.1103/PhysRevD.80.065021} {\bibfield  {journal}
  {\bibinfo  {journal} {Phys. Rev.}\ }\textbf {\bibinfo {volume} {D80}},\
  \bibinfo {pages} {065021} (\bibinfo {year} {2009})},\ \Eprint
  {http://arxiv.org/abs/0903.3946} {arXiv:0903.3946 [hep-ph]} \BibitemShut
  {NoStop}%
\bibitem [{\citenamefont {Baier}(2009)}]{Baier:2009zy}%
  \BibitemOpen
  \bibfield  {author} {\bibinfo {author} {\bibfnamefont {R.}~\bibnamefont
  {Baier}},\ }\href@noop {} {\  (\bibinfo {year} {2009})},\ \Eprint
  {http://arxiv.org/abs/0910.3862} {arXiv:0910.3862 [hep-th]} \BibitemShut
  {NoStop}%
\bibitem [{\citenamefont {Meyer}(2010{\natexlab{a}})}]{Meyer:2010ii}%
  \BibitemOpen
  \bibfield  {author} {\bibinfo {author} {\bibfnamefont {H.~B.}\ \bibnamefont
  {Meyer}},\ }\href {\doibase 10.1007/JHEP04(2010)099} {\bibfield  {journal}
  {\bibinfo  {journal} {JHEP}\ }\textbf {\bibinfo {volume} {04}},\ \bibinfo
  {pages} {099} (\bibinfo {year} {2010}{\natexlab{a}})},\ \Eprint
  {http://arxiv.org/abs/1002.3343} {arXiv:1002.3343 [hep-lat]} \BibitemShut
  {NoStop}%
\bibitem [{\citenamefont {Meyer}(2010{\natexlab{b}})}]{Meyer:2010gu}%
  \BibitemOpen
  \bibfield  {author} {\bibinfo {author} {\bibfnamefont {H.~B.}\ \bibnamefont
  {Meyer}},\ }\href {\doibase 10.1103/PhysRevD.82.054504} {\bibfield  {journal}
  {\bibinfo  {journal} {Phys.Rev.}\ }\textbf {\bibinfo {volume} {D82}},\
  \bibinfo {pages} {054504} (\bibinfo {year} {2010}{\natexlab{b}})},\ \Eprint
  {http://arxiv.org/abs/arXiv:1005.2686} {arXiv:arXiv:1005.2686 [hep-lat]}
  \BibitemShut {NoStop}%
\bibitem [{\citenamefont {Maldacena}(1998)}]{Maldacena:1997re}%
  \BibitemOpen
  \bibfield  {author} {\bibinfo {author} {\bibfnamefont {J.~M.}\ \bibnamefont
  {Maldacena}},\ }\href@noop {} {\bibfield  {journal} {\bibinfo  {journal}
  {Adv. Theor. Math. Phys.}\ }\textbf {\bibinfo {volume} {2}},\ \bibinfo
  {pages} {231} (\bibinfo {year} {1998})},\ \Eprint
  {http://arxiv.org/abs/hep-th/9711200} {arXiv:hep-th/9711200} \BibitemShut
  {NoStop}%
\bibitem [{\citenamefont {Witten}(1998)}]{Witten:1998qj}%
  \BibitemOpen
  \bibfield  {author} {\bibinfo {author} {\bibfnamefont {E.}~\bibnamefont
  {Witten}},\ }\href@noop {} {\bibfield  {journal} {\bibinfo  {journal} {Adv.
  Theor. Math. Phys.}\ }\textbf {\bibinfo {volume} {2}},\ \bibinfo {pages}
  {253} (\bibinfo {year} {1998})},\ \Eprint
  {http://arxiv.org/abs/hep-th/9802150} {arXiv:hep-th/9802150} \BibitemShut
  {NoStop}%
\bibitem [{\citenamefont {Gubser}\ \emph {et~al.}(1998)\citenamefont {Gubser},
  \citenamefont {Klebanov},\ and\ \citenamefont {Polyakov}}]{Gubser:1998bc}%
  \BibitemOpen
  \bibfield  {author} {\bibinfo {author} {\bibfnamefont {S.~S.}\ \bibnamefont
  {Gubser}}, \bibinfo {author} {\bibfnamefont {I.~R.}\ \bibnamefont
  {Klebanov}}, \ and\ \bibinfo {author} {\bibfnamefont {A.~M.}\ \bibnamefont
  {Polyakov}},\ }\href {\doibase 10.1016/S0370-2693(98)00377-3} {\bibfield
  {journal} {\bibinfo  {journal} {Phys. Lett.}\ }\textbf {\bibinfo {volume}
  {B428}},\ \bibinfo {pages} {105} (\bibinfo {year} {1998})},\ \Eprint
  {http://arxiv.org/abs/hep-th/9802109} {arXiv:hep-th/9802109} \BibitemShut
  {NoStop}%
\bibitem [{\citenamefont {Son}\ and\ \citenamefont
  {Starinets}(2002)}]{Son:2002sd}%
  \BibitemOpen
  \bibfield  {author} {\bibinfo {author} {\bibfnamefont {D.~T.}\ \bibnamefont
  {Son}}\ and\ \bibinfo {author} {\bibfnamefont {A.~O.}\ \bibnamefont
  {Starinets}},\ }\href@noop {} {\bibfield  {journal} {\bibinfo  {journal}
  {JHEP}\ }\textbf {\bibinfo {volume} {09}},\ \bibinfo {pages} {042} (\bibinfo
  {year} {2002})},\ \Eprint {http://arxiv.org/abs/hep-th/0205051}
  {arXiv:hep-th/0205051} \BibitemShut {NoStop}%
\bibitem [{\citenamefont {Myers}\ \emph {et~al.}(2007)\citenamefont {Myers},
  \citenamefont {Starinets},\ and\ \citenamefont {Thomson}}]{Myers:2007we}%
  \BibitemOpen
  \bibfield  {author} {\bibinfo {author} {\bibfnamefont {R.~C.}\ \bibnamefont
  {Myers}}, \bibinfo {author} {\bibfnamefont {A.~O.}\ \bibnamefont
  {Starinets}}, \ and\ \bibinfo {author} {\bibfnamefont {R.~M.}\ \bibnamefont
  {Thomson}},\ }\href {\doibase 10.1088/1126-6708/2007/11/091} {\bibfield
  {journal} {\bibinfo  {journal} {JHEP}\ }\textbf {\bibinfo {volume} {11}},\
  \bibinfo {pages} {091} (\bibinfo {year} {2007})},\ \Eprint
  {http://arxiv.org/abs/0706.0162} {arXiv:0706.0162 [hep-th]} \BibitemShut
  {NoStop}%
\bibitem [{\citenamefont {Chamblin}\ and\ \citenamefont
  {Reall}(1999)}]{Chamblin:1999ya}%
  \BibitemOpen
  \bibfield  {author} {\bibinfo {author} {\bibfnamefont {H.~A.}\ \bibnamefont
  {Chamblin}}\ and\ \bibinfo {author} {\bibfnamefont {H.~S.}\ \bibnamefont
  {Reall}},\ }\href {\doibase 10.1016/S0550-3213(99)00520-9} {\bibfield
  {journal} {\bibinfo  {journal} {Nucl. Phys.}\ }\textbf {\bibinfo {volume}
  {B562}},\ \bibinfo {pages} {133} (\bibinfo {year} {1999})},\ \Eprint
  {http://arxiv.org/abs/hep-th/9903225} {arXiv:hep-th/9903225} \BibitemShut
  {NoStop}%
\bibitem [{\citenamefont {Springer}(2009{\natexlab{a}})}]{Springer:2008js}%
  \BibitemOpen
  \bibfield  {author} {\bibinfo {author} {\bibfnamefont {T.}~\bibnamefont
  {Springer}},\ }\href {\doibase 10.1103/PhysRevD.79.046003} {\bibfield
  {journal} {\bibinfo  {journal} {Phys. Rev.}\ }\textbf {\bibinfo {volume}
  {D79}},\ \bibinfo {pages} {046003} (\bibinfo {year} {2009}{\natexlab{a}})},\
  \Eprint {http://arxiv.org/abs/0810.4354} {arXiv:0810.4354 [hep-th]}
  \BibitemShut {NoStop}%
\bibitem [{\citenamefont {Springer}(2009{\natexlab{b}})}]{Springer:2009wj}%
  \BibitemOpen
  \bibfield  {author} {\bibinfo {author} {\bibfnamefont {T.}~\bibnamefont
  {Springer}},\ }\href {\doibase 10.1103/PhysRevD.79.086003} {\bibfield
  {journal} {\bibinfo  {journal} {Phys. Rev.}\ }\textbf {\bibinfo {volume}
  {D79}},\ \bibinfo {pages} {086003} (\bibinfo {year} {2009}{\natexlab{b}})},\
  \Eprint {http://arxiv.org/abs/0902.2566} {arXiv:0902.2566 [hep-th]}
  \BibitemShut {NoStop}%
\bibitem [{\citenamefont {Gubser}\ and\ \citenamefont
  {Nellore}(2008)}]{Gubser:2008ny}%
  \BibitemOpen
  \bibfield  {author} {\bibinfo {author} {\bibfnamefont {S.~S.}\ \bibnamefont
  {Gubser}}\ and\ \bibinfo {author} {\bibfnamefont {A.}~\bibnamefont
  {Nellore}},\ }\href {\doibase 10.1103/PhysRevD.78.086007} {\bibfield
  {journal} {\bibinfo  {journal} {Phys. Rev.}\ }\textbf {\bibinfo {volume}
  {D78}},\ \bibinfo {pages} {086007} (\bibinfo {year} {2008})},\ \Eprint
  {http://arxiv.org/abs/0804.0434} {arXiv:0804.0434 [hep-th]} \BibitemShut
  {NoStop}%
\bibitem [{\citenamefont {Gubser}\ \emph
  {et~al.}(2008{\natexlab{a}})\citenamefont {Gubser}, \citenamefont {Nellore},
  \citenamefont {Pufu},\ and\ \citenamefont {Rocha}}]{Gubser:2008yx}%
  \BibitemOpen
  \bibfield  {author} {\bibinfo {author} {\bibfnamefont {S.~S.}\ \bibnamefont
  {Gubser}}, \bibinfo {author} {\bibfnamefont {A.}~\bibnamefont {Nellore}},
  \bibinfo {author} {\bibfnamefont {S.~S.}\ \bibnamefont {Pufu}}, \ and\
  \bibinfo {author} {\bibfnamefont {F.~D.}\ \bibnamefont {Rocha}},\ }\href
  {\doibase 10.1103/PhysRevLett.101.131601} {\bibfield  {journal} {\bibinfo
  {journal} {Phys. Rev. Lett.}\ }\textbf {\bibinfo {volume} {101}},\ \bibinfo
  {pages} {131601} (\bibinfo {year} {2008}{\natexlab{a}})},\ \Eprint
  {http://arxiv.org/abs/0804.1950} {arXiv:0804.1950 [hep-th]} \BibitemShut
  {NoStop}%
\bibitem [{\citenamefont {Kanitscheider}\ \emph {et~al.}(2008)\citenamefont
  {Kanitscheider}, \citenamefont {Skenderis},\ and\ \citenamefont
  {Taylor}}]{Kanitscheider:2008kd}%
  \BibitemOpen
  \bibfield  {author} {\bibinfo {author} {\bibfnamefont {I.}~\bibnamefont
  {Kanitscheider}}, \bibinfo {author} {\bibfnamefont {K.}~\bibnamefont
  {Skenderis}}, \ and\ \bibinfo {author} {\bibfnamefont {M.}~\bibnamefont
  {Taylor}},\ }\href {\doibase 10.1088/1126-6708/2008/09/094} {\bibfield
  {journal} {\bibinfo  {journal} {JHEP}\ }\textbf {\bibinfo {volume} {09}},\
  \bibinfo {pages} {094} (\bibinfo {year} {2008})},\ \Eprint
  {http://arxiv.org/abs/0807.3324} {arXiv:0807.3324 [hep-th]} \BibitemShut
  {NoStop}%
\bibitem [{\citenamefont {Romatschke}(2010)}]{Romatschke:2009kr}%
  \BibitemOpen
  \bibfield  {author} {\bibinfo {author} {\bibfnamefont {P.}~\bibnamefont
  {Romatschke}},\ }\href {\doibase 10.1088/0264-9381/27/2/025006} {\bibfield
  {journal} {\bibinfo  {journal} {Class. Quant. Grav.}\ }\textbf {\bibinfo
  {volume} {27}},\ \bibinfo {pages} {025006} (\bibinfo {year} {2010})},\
  \Eprint {http://arxiv.org/abs/0906.4787} {arXiv:0906.4787 [hep-th]}
  \BibitemShut {NoStop}%
\bibitem [{\citenamefont {Bigazzi}\ and\ \citenamefont
  {Cotrone}(2010)}]{Bigazzi:2010ku}%
  \BibitemOpen
  \bibfield  {author} {\bibinfo {author} {\bibfnamefont {F.}~\bibnamefont
  {Bigazzi}}\ and\ \bibinfo {author} {\bibfnamefont {A.~L.}\ \bibnamefont
  {Cotrone}},\ }\href {\doibase 10.1007/JHEP08(2010)128} {\bibfield  {journal}
  {\bibinfo  {journal} {JHEP}\ }\textbf {\bibinfo {volume} {1008}},\ \bibinfo
  {pages} {128} (\bibinfo {year} {2010})},\ \Eprint
  {http://arxiv.org/abs/arXiv:1006.4634} {arXiv:arXiv:1006.4634 [hep-ph]}
  \BibitemShut {NoStop}%
\bibitem [{\citenamefont {Bigazzi}\ \emph {et~al.}(2010)\citenamefont
  {Bigazzi}, \citenamefont {Cotrone},\ and\ \citenamefont
  {Tarrio}}]{Bigazzi:2009tc}%
  \BibitemOpen
  \bibfield  {author} {\bibinfo {author} {\bibfnamefont {F.}~\bibnamefont
  {Bigazzi}}, \bibinfo {author} {\bibfnamefont {A.~L.}\ \bibnamefont
  {Cotrone}}, \ and\ \bibinfo {author} {\bibfnamefont {J.}~\bibnamefont
  {Tarrio}},\ }\href {\doibase 10.1007/JHEP02(2010)083} {\bibfield  {journal}
  {\bibinfo  {journal} {JHEP}\ }\textbf {\bibinfo {volume} {02}},\ \bibinfo
  {pages} {083} (\bibinfo {year} {2010})},\ \Eprint
  {http://arxiv.org/abs/0912.3256} {arXiv:0912.3256 [hep-th]} \BibitemShut
  {NoStop}%
\bibitem [{\citenamefont {Kanitscheider}\ and\ \citenamefont
  {Skenderis}(2009)}]{Kanitscheider:2009as}%
  \BibitemOpen
  \bibfield  {author} {\bibinfo {author} {\bibfnamefont {I.}~\bibnamefont
  {Kanitscheider}}\ and\ \bibinfo {author} {\bibfnamefont {K.}~\bibnamefont
  {Skenderis}},\ }\href {\doibase 10.1088/1126-6708/2009/04/062} {\bibfield
  {journal} {\bibinfo  {journal} {JHEP}\ }\textbf {\bibinfo {volume} {04}},\
  \bibinfo {pages} {062} (\bibinfo {year} {2009})},\ \Eprint
  {http://arxiv.org/abs/0901.1487} {arXiv:0901.1487 [hep-th]} \BibitemShut
  {NoStop}%
\bibitem [{\citenamefont {Buchel}\ \emph {et~al.}(2005)\citenamefont {Buchel},
  \citenamefont {Liu},\ and\ \citenamefont {Starinets}}]{Buchel:2004di}%
  \BibitemOpen
  \bibfield  {author} {\bibinfo {author} {\bibfnamefont {A.}~\bibnamefont
  {Buchel}}, \bibinfo {author} {\bibfnamefont {J.~T.}\ \bibnamefont {Liu}}, \
  and\ \bibinfo {author} {\bibfnamefont {A.~O.}\ \bibnamefont {Starinets}},\
  }\href {\doibase 10.1016/j.nuclphysb.2004.11.055} {\bibfield  {journal}
  {\bibinfo  {journal} {Nucl. Phys.}\ }\textbf {\bibinfo {volume} {B707}},\
  \bibinfo {pages} {56} (\bibinfo {year} {2005})},\ \Eprint
  {http://arxiv.org/abs/hep-th/0406264} {arXiv:hep-th/0406264} \BibitemShut
  {NoStop}%
\bibitem [{\citenamefont {Gubser}\ \emph
  {et~al.}(2008{\natexlab{b}})\citenamefont {Gubser}, \citenamefont {Pufu},\
  and\ \citenamefont {Rocha}}]{Gubser:2008sz}%
  \BibitemOpen
  \bibfield  {author} {\bibinfo {author} {\bibfnamefont {S.~S.}\ \bibnamefont
  {Gubser}}, \bibinfo {author} {\bibfnamefont {S.~S.}\ \bibnamefont {Pufu}}, \
  and\ \bibinfo {author} {\bibfnamefont {F.~D.}\ \bibnamefont {Rocha}},\ }\href
  {\doibase 10.1088/1126-6708/2008/08/085} {\bibfield  {journal} {\bibinfo
  {journal} {JHEP}\ }\textbf {\bibinfo {volume} {08}},\ \bibinfo {pages} {085}
  (\bibinfo {year} {2008}{\natexlab{b}})},\ \Eprint
  {http://arxiv.org/abs/0806.0407} {arXiv:0806.0407 [hep-th]} \BibitemShut
  {NoStop}%
\bibitem [{\citenamefont {Kaminski}\ \emph {et~al.}(2010)\citenamefont
  {Kaminski}, \citenamefont {Landsteiner}, \citenamefont {Mas}, \citenamefont
  {Shock},\ and\ \citenamefont {Tarrio}}]{Kaminski:2009dh}%
  \BibitemOpen
  \bibfield  {author} {\bibinfo {author} {\bibfnamefont {M.}~\bibnamefont
  {Kaminski}}, \bibinfo {author} {\bibfnamefont {K.}~\bibnamefont
  {Landsteiner}}, \bibinfo {author} {\bibfnamefont {J.}~\bibnamefont {Mas}},
  \bibinfo {author} {\bibfnamefont {J.~P.}\ \bibnamefont {Shock}}, \ and\
  \bibinfo {author} {\bibfnamefont {J.}~\bibnamefont {Tarrio}},\ }\href
  {\doibase 10.1007/JHEP02(2010)021} {\bibfield  {journal} {\bibinfo  {journal}
  {JHEP}\ }\textbf {\bibinfo {volume} {02}},\ \bibinfo {pages} {021} (\bibinfo
  {year} {2010})},\ \Eprint {http://arxiv.org/abs/0911.3610} {arXiv:0911.3610
  [hep-th]} \BibitemShut {NoStop}%
\bibitem [{Mat(2008)}]{Mathematica}%
  \BibitemOpen
  \href@noop {} {\emph {\bibinfo {title} {Mathematica 7.0}}}\ (\bibinfo
  {publisher} {Wolfram Research, Inc., Champaign, IL},\ \bibinfo {year}
  {2008})\BibitemShut {NoStop}%
\bibitem [{\citenamefont {Kovtun}\ and\ \citenamefont
  {Starinets}(2006)}]{Kovtun:2006pf}%
  \BibitemOpen
  \bibfield  {author} {\bibinfo {author} {\bibfnamefont {P.}~\bibnamefont
  {Kovtun}}\ and\ \bibinfo {author} {\bibfnamefont {A.}~\bibnamefont
  {Starinets}},\ }\href {\doibase 10.1103/PhysRevLett.96.131601} {\bibfield
  {journal} {\bibinfo  {journal} {Phys. Rev. Lett.}\ }\textbf {\bibinfo
  {volume} {96}},\ \bibinfo {pages} {131601} (\bibinfo {year} {2006})},\
  \Eprint {http://arxiv.org/abs/hep-th/0602059} {arXiv:hep-th/0602059}
  \BibitemShut {NoStop}%
\bibitem [{\citenamefont {Teaney}(2006)}]{Teaney:2006nc}%
  \BibitemOpen
  \bibfield  {author} {\bibinfo {author} {\bibfnamefont {D.}~\bibnamefont
  {Teaney}},\ }\href {\doibase 10.1103/PhysRevD.74.045025} {\bibfield
  {journal} {\bibinfo  {journal} {Phys. Rev.}\ }\textbf {\bibinfo {volume}
  {D74}},\ \bibinfo {pages} {045025} (\bibinfo {year} {2006})},\ \Eprint
  {http://arxiv.org/abs/hep-ph/0602044} {arXiv:hep-ph/0602044} \BibitemShut
  {NoStop}%
\bibitem [{\citenamefont {Hartnoll}\ and\ \citenamefont
  {Prem~Kumar}(2005)}]{Hartnoll:2005ju}%
  \BibitemOpen
  \bibfield  {author} {\bibinfo {author} {\bibfnamefont {S.~A.}\ \bibnamefont
  {Hartnoll}}\ and\ \bibinfo {author} {\bibfnamefont {S.}~\bibnamefont
  {Prem~Kumar}},\ }\href@noop {} {\bibfield  {journal} {\bibinfo  {journal}
  {JHEP}\ }\textbf {\bibinfo {volume} {12}},\ \bibinfo {pages} {036} (\bibinfo
  {year} {2005})},\ \Eprint {http://arxiv.org/abs/hep-th/0508092}
  {arXiv:hep-th/0508092} \BibitemShut {NoStop}%
\bibitem [{\citenamefont {Nakamura}\ and\ \citenamefont
  {Sakai}(2005)}]{Nakamura:2004sy}%
  \BibitemOpen
  \bibfield  {author} {\bibinfo {author} {\bibfnamefont {A.}~\bibnamefont
  {Nakamura}}\ and\ \bibinfo {author} {\bibfnamefont {S.}~\bibnamefont
  {Sakai}},\ }\href {\doibase 10.1103/PhysRevLett.94.072305} {\bibfield
  {journal} {\bibinfo  {journal} {Phys. Rev. Lett.}\ }\textbf {\bibinfo
  {volume} {94}},\ \bibinfo {pages} {072305} (\bibinfo {year} {2005})},\
  \Eprint {http://arxiv.org/abs/hep-lat/0406009} {arXiv:hep-lat/0406009}
  \BibitemShut {NoStop}%
\bibitem [{\citenamefont {Meyer}(2007)}]{Meyer:2007ic}%
  \BibitemOpen
  \bibfield  {author} {\bibinfo {author} {\bibfnamefont {H.~B.}\ \bibnamefont
  {Meyer}},\ }\href {\doibase 10.1103/PhysRevD.76.101701} {\bibfield  {journal}
  {\bibinfo  {journal} {Phys. Rev.}\ }\textbf {\bibinfo {volume} {D76}},\
  \bibinfo {pages} {101701} (\bibinfo {year} {2007})},\ \Eprint
  {http://arxiv.org/abs/0704.1801} {arXiv:0704.1801 [hep-lat]} \BibitemShut
  {NoStop}%
\bibitem [{\citenamefont {Gursoy}\ \emph {et~al.}(2009)\citenamefont {Gursoy},
  \citenamefont {Kiritsis}, \citenamefont {Mazzanti},\ and\ \citenamefont
  {Nitti}}]{Gursoy:2008za}%
  \BibitemOpen
  \bibfield  {author} {\bibinfo {author} {\bibfnamefont {U.}~\bibnamefont
  {Gursoy}}, \bibinfo {author} {\bibfnamefont {E.}~\bibnamefont {Kiritsis}},
  \bibinfo {author} {\bibfnamefont {L.}~\bibnamefont {Mazzanti}}, \ and\
  \bibinfo {author} {\bibfnamefont {F.}~\bibnamefont {Nitti}},\ }\href
  {\doibase 10.1088/1126-6708/2009/05/033} {\bibfield  {journal} {\bibinfo
  {journal} {JHEP}\ }\textbf {\bibinfo {volume} {05}},\ \bibinfo {pages} {033}
  (\bibinfo {year} {2009})},\ \Eprint {http://arxiv.org/abs/0812.0792}
  {arXiv:0812.0792 [hep-th]} \BibitemShut {NoStop}%
\bibitem [{\citenamefont {Gursoy}\ \emph {et~al.}(2008)\citenamefont {Gursoy},
  \citenamefont {Kiritsis}, \citenamefont {Mazzanti},\ and\ \citenamefont
  {Nitti}}]{Gursoy:2008bu}%
  \BibitemOpen
  \bibfield  {author} {\bibinfo {author} {\bibfnamefont {U.}~\bibnamefont
  {Gursoy}}, \bibinfo {author} {\bibfnamefont {E.}~\bibnamefont {Kiritsis}},
  \bibinfo {author} {\bibfnamefont {L.}~\bibnamefont {Mazzanti}}, \ and\
  \bibinfo {author} {\bibfnamefont {F.}~\bibnamefont {Nitti}},\ }\href
  {\doibase 10.1103/PhysRevLett.101.181601} {\bibfield  {journal} {\bibinfo
  {journal} {Phys. Rev. Lett.}\ }\textbf {\bibinfo {volume} {101}},\ \bibinfo
  {pages} {181601} (\bibinfo {year} {2008})},\ \Eprint
  {http://arxiv.org/abs/0804.0899} {arXiv:0804.0899 [hep-th]} \BibitemShut
  {NoStop}%
\bibitem [{\citenamefont {Caron-Huot}(2009)}]{CaronHuot:2009ns}%
  \BibitemOpen
  \bibfield  {author} {\bibinfo {author} {\bibfnamefont {S.}~\bibnamefont
  {Caron-Huot}},\ }\href {\doibase 10.1103/PhysRevD.79.125009} {\bibfield
  {journal} {\bibinfo  {journal} {Phys. Rev.}\ }\textbf {\bibinfo {volume}
  {D79}},\ \bibinfo {pages} {125009} (\bibinfo {year} {2009})},\ \Eprint
  {http://arxiv.org/abs/0903.3958} {arXiv:0903.3958 [hep-ph]} \BibitemShut
  {NoStop}%
\bibitem [{\citenamefont {Lee}\ and\ \citenamefont
  {Morita}(2009)}]{Lee:2008xp}%
  \BibitemOpen
  \bibfield  {author} {\bibinfo {author} {\bibfnamefont {S.~H.}\ \bibnamefont
  {Lee}}\ and\ \bibinfo {author} {\bibfnamefont {K.}~\bibnamefont {Morita}},\
  }\href {\doibase 10.1103/PhysRevD.79.011501} {\bibfield  {journal} {\bibinfo
  {journal} {Phys. Rev.}\ }\textbf {\bibinfo {volume} {D79}},\ \bibinfo {pages}
  {011501} (\bibinfo {year} {2009})},\ \Eprint {http://arxiv.org/abs/0802.4000}
  {arXiv:0802.4000 [hep-ph]} \BibitemShut {NoStop}%
\bibitem [{\citenamefont {Kajantie}\ \emph {et~al.}(2007)\citenamefont
  {Kajantie}, \citenamefont {Tahkokallio},\ and\ \citenamefont
  {Yee}}]{Kajantie:2006hv}%
  \BibitemOpen
  \bibfield  {author} {\bibinfo {author} {\bibfnamefont {K.}~\bibnamefont
  {Kajantie}}, \bibinfo {author} {\bibfnamefont {T.}~\bibnamefont
  {Tahkokallio}}, \ and\ \bibinfo {author} {\bibfnamefont {J.-T.}\ \bibnamefont
  {Yee}},\ }\href@noop {} {\bibfield  {journal} {\bibinfo  {journal} {JHEP}\
  }\textbf {\bibinfo {volume} {01}},\ \bibinfo {pages} {019} (\bibinfo {year}
  {2007})},\ \Eprint {http://arxiv.org/abs/hep-ph/0609254}
  {arXiv:hep-ph/0609254} \BibitemShut {NoStop}%
\end{thebibliography}%

\end{document}